\documentclass[prd,amsmath,superscriptaddress,twocolumn,nofootinbib]{revtex4-2}
\pdfoutput=1

\usepackage{textcomp}
\usepackage[utf8]{inputenc}
\usepackage{graphicx}
\usepackage{float}
\usepackage{epsf,latexsym,bbm,euscript}
\usepackage{amssymb,amsmath}
\usepackage{mathtools} 
\usepackage{times}
\usepackage{soul}
\usepackage{mathtools}
\usepackage{mathrsfs}
\usepackage{array}
\usepackage{booktabs}
\usepackage{enumitem}
\usepackage{tensor}
\usepackage[dvipsnames]{xcolor}
\usepackage{tikz}
\usetikzlibrary{calc}

\usepackage[caption=false]{subfig}
\newcommand{\RN}[1]{\uppercase\expandafter{\romannumeral #1\relax}}

\usepackage{environ}
\NewEnviron{thincases}{\scalebox{0.5}[1]{$\displaystyle
		\left\{\scalebox{2}[1]{\setlength{\arraycolsep}{6pt}
			$\displaystyle\begin{array}{ll}
				\BODY
			\end{array}$}\right.$}}%}
			\NewEnviron{thinpmatrix}{\scalebox{0.5}[1]{$\displaystyle
	\left(\scalebox{2}[1]{$\displaystyle
		\begin{matrix}
			\BODY
		\end{matrix}$}
	\right)$}}
	
	\makeatletter
	\g@addto@macro\bfseries{\boldmath}
	\makeatother
	
	\makeatletter
	\newcommand*{\defeq}{\mathrel{\rlap{%
		\raisebox{0.3ex}{$\m@th\cdot$}}%
	\raisebox{-0.3ex}{$\m@th\cdot$}}%
=}
\newcommand*{\eqdef}{=\mathrel{\rlap{%
		\raisebox{0.3ex}{$\m@th\cdot$}}%
	\raisebox{-0.3ex}{$\m@th\cdot$}}%
	}
	\makeatother

	\usepackage{pifont}
	\newcommand{\cmark}{\text{\ding{51}}}
	\newcommand{\xmark}{\text{\ding{55}}}
	
	\usepackage{scalerel}
	\usepackage{tikz}
	\usetikzlibrary{svg.path}
	\definecolor{orcidlogocol}{HTML}{A6CE39}
	\tikzset{
orcidlogo/.pic={
	\fill[orcidlogocol] svg{M256,128c0,70.7-57.3,128-128,128C57.3,256,0,198.7,0,128C0,57.3,57.3,0,128,0C198.7,0,256,57.3,256,128z};
	\fill[white] svg{M86.3,186.2H70.9V79.1h15.4v48.4V186.2z}
	svg{M108.9,79.1h41.6c39.6,0,57,28.3,57,53.6c0,27.5-21.5,53.6-56.8,53.6h-41.8V79.1z M124.3,172.4h24.5c34.9,0,42.9-26.5,42.9-39.7c0-21.5-13.7-39.7-43.7-39.7h-23.7V172.4z}
	svg{M88.7,56.8c0,5.5-4.5,10.1-10.1,10.1c-5.6,0-10.1-4.6-10.1-10.1c0-5.6,4.5-10.1,10.1-10.1C84.2,46.7,88.7,51.3,88.7,56.8z};
}
}
\newcommand\orcidlink[1]{\href{https://orcid.org/#1}{\mbox{\scalerel*{
			\begin{tikzpicture}[yscale=-1,transform shape]
				\pic{orcidlogo};
			\end{tikzpicture}
		}{X}}}}
		
		\usepackage{url,hyperref}
		\hypersetup{colorlinks,linkcolor={blue!55!black},citecolor={red!50!black},urlcolor={blue!45!black},breaklinks=true}
		
		\usetikzlibrary{arrows.meta,calc,decorations.markings}

		\newcommand{\mA}{\mathrm{A}}
			\newcommand{\mB}{\mathrm{B}}
				\newcommand{\mC}{\mathrm{C}}
					\newcommand{\mD}{\mathrm{D}}
		
		\usepackage{url,hyperref}
		\hypersetup{colorlinks,linkcolor={blue!55!black},citecolor={red!50!black},urlcolor={blue!45!black}}

\usepackage[colorinlistoftodos,prependcaption,textsize=tiny,loadshadowlibrary]{todonotes}

\begin{document}

\title{Probing mass inflation in polymerized vacuum regular black holes\\ via colliding null shells}
\author{Hongguang Liu\orcidlink{0000-0002-1059-045X}}
\email{liuhongguang@westlake.edu.cn}

\author{Ioannis Soranidis\orcidlink{0000-0002-8652-9874}}
\email{ioannis.soranidis@westlake.edu.cn}

\affiliation{Institute for Theoretical Sciences, and Departments of Physics and Astronomy, Westlake University, Hangzhou 310030, China}
\affiliation{Institute of Natural Sciences, Westlake Institute for Advanced Study, Hangzhou 310024, China}

\begin{abstract}
\vspace*{0.2cm}	
We derive a class of inner-extremal regular black hole solutions characterized by a degenerate inner horizon. These geometries arise as polymerized vacuum configurations inspired by loop quantum gravity and constitute effective quantum-gravity solutions that admit a Birkhoff-type theorem, rendering them unique within the considered framework. We show that such inner-extremal horizon configurations exist only for a finely tuned value of the mass determined by the parameters of the theory. Building on this construction, together with the corresponding non-degenerate regular black hole solutions, we perform a generic analysis of the mass inflation phenomenon in four-dimensional spacetimes using a colliding null-shell setup near the inner horizon. We identify the conditions under which mass inflation becomes significant and examine how the presence of a minimal length scale affects this behavior, with particular emphasis on the case where such a scale is motivated by loop quantum gravity. Finally, we comment on the stability of these configurations under the null-shell perturbations considered in our analysis.

\end{abstract}
\maketitle

\section{introduction}\label{sec:intro}

A fundamental expectation from any consistent theory of quantum gravity is the removal of spacetime singularities that arise in cosmological settings and in the interior of black holes. In classical general relativity, such singularities are associated with divergent curvature invariants and signal the breakdown of the theory’s predictive power \cite{H:76}. At the same time, observations provide strong evidence for the existence of dark, massive objects that are sufficiently compact to sustain light rings, placing them in the class of ultracompact objects. While their presence in the Universe is well established, their precise nature remains an open question \cite{CP:19, BCNS:19, M:23}. This uncertainty has motivated a variety of theoretical proposals, many of which rely on effective descriptions designed to incorporate singularity-resolution mechanisms.

Representative examples include regular black holes (RBHs) \cite{B:68, D:92, H:06}, gravastars \cite{MM:04,MM:23-grav}, wormholes \cite{E:73,MT:88,SV:19}, and fuzzballs \cite{LM:02, M:05}. These proposals achieve singularity resolution through different mechanisms. In some cases, the event horizon is entirely absent, resulting in horizonless configurations. In others, the classical singularity is effectively replaced by an inner horizon, a feature often interpreted as emerging from quantum-gravity corrections that introduce a fundamental minimal length scale. Nevertheless, this seemingly benign replacement can give rise to nontrivial complications in certain physical scenarios.

A paradigmatic example illustrating the subtleties associated with inner horizons is the Reissner–Nordström spacetime. In this case, a nonzero subextremal electric charge, $0<|Q|<M$, gives rise to a two-horizon structure, consisting of an outer event horizon and an inner Cauchy horizon. The inner horizon introduces a region of extreme blueshift, rendering it highly sensitive to perturbations. This leads to the well-known mass inflation scenario, where arbitrarily small disturbances can trigger large curvature growth in the vicinity of the Cauchy horizon\footnote{The origin of the term “mass inflation” lies in the rapid growth of the Misner–Sharp mass near the inner horizon of charged black holes.} \cite{PI:89, PI:90, BS:95}. This phenomenon has been studied in great detail for various scenarios \cite{HA:10,HY:11}. It is worth stressing that an actual divergence of curvature invariants is not necessary to diagnose a physical instability. Even when all geometric quantities remain finite, an exponential amplification of energy densities and curvature can drive them to parametrically large values. Once such scales are reached, the classical description can no longer be considered reliable \cite{CRFLV:24}. 

RBHs generally exhibit a similar multi-horizon structure, consisting of an outer event horizon and an inner horizon that replaces the classical singularity. This naturally raises the question of whether comparable dynamical phenomena may occur in such geometries. In stationary spacetimes, the inner horizon of RBHs is necessarily a Cauchy horizon. If it fulfills a role analogous to that of the Cauchy horizon in the Reissner–Nordström solution, then its stability under perturbations becomes a fundamental requirement for the internal consistency of the model. Addressing this question is essential for assessing whether the effective resolution of singularities truly leads to a physically viable spacetime. Numerous studies have investigated this issue in the context of RBHs \cite{FZ:26,FZ:17,BMM:11,BKS:21,HKSWE:22,CRFLPV:22,BRSZ:21,BF:22,Bambi:book:23,BBCRG:22,Mc:23,BPS:25,CLWZ:24}, consistently indicating that their inner horizons are generically susceptible to mass inflation, although in certain scenarios the effect may be partially mitigated, potentially preventing the formation of a singularity. One possible way to mitigate this instability, at least at the classical level\footnote{We should note that once semiclassical effects are incorporated, the inner-horizon instability persists for generic inner-extremal RBHs formed through gravitational collapse \cite{Mc:23}.}, is to consider configurations in which the inner horizon becomes degenerate \cite{CRFLCV:22:deg}.

Regular black hole solutions can be constructed through a variety of mechanisms, both with and without matter sources. A well-known example of the former arises in nonlinear electrodynamics, where a magnetic monopole configuration regularizes the spacetime within general relativity \cite{pt:69,bmss:79,ag:98,ag:99a,ag:99b,b:00,b:01,bh:02,d:04,bv:14,fw:16,b:17,tsa:18,b:23}. More recently, however, it has been shown that vacuum regular black hole solutions can also be obtained. Such constructions appear in quasitopological gravity for dimensions $D\geqslant 5$ \cite{BCH:25,BCHM:25a,BCHM:25b,BCHMC:25,HKMS:25,FKSZ:25}, and analogous vacuum configurations in $D=4$ have been derived within alternative frameworks \cite{BenAchour:2018khr,Han:2020uhb,Han:2022rsx,F:25,BCR:26,BCHM:26,GLRSW:25,GLSW:25,GL:25}.\footnote{A recent general reconstruction result demonstrates that a broad class of static, spherically symmetric, asymptotically flat geometries satisfying $g_{tt}g_{rr}=-1$, including regular black-hole geometries, can be realized as vacuum solutions of extended quasitopological gravities in $D\geqslant4$~\cite{B:26}. Related four-dimensional nonpolynomial quasitopological theories have also been applied to cosmology, where appropriate choices of the characteristic function yield nonsingular FLRW evolutions~\cite{BM:26}.}

The polymerized-vacuum construction underlying our analysis is reviewed in detail and implemented for anti-de Sitter-core ultracompact objects in our work~\cite{LS:26}, building on Refs.~\cite{GLSW:25,GL:25}. The construction starts from a dust-time-deparameterized Lemaître--Tolman--Bondi (LTB) description of spherically symmetric collapse and exploits the decoupling of the shell dynamics. On the vacuum branch, the dust density is set to zero, so the dust serves only as an auxiliary relational clock and foliation variable and does not source the geometry. Requiring a static vacuum geometry independent of the LTB energy function leads to Eqs.~\eqref{eq:ds} and~\eqref{eq:f}. The model-defining function $\tilde f$, subject to the Schwarzschild low-curvature limit and the regularity conditions of the construction\footnote{A detailed characterization of the conditions that $\tilde f$ must satisfy for the resulting geometry to describe a regular black hole is provided in Ref.~\cite{BS:26}.}, specifies the effective polymerized theory; once $\tilde f$ and the vacuum branch are fixed, the Birkhoff-type property implies that the static solution is uniquely characterized by the conserved integration constant $m$. Ref.~\cite{LS:26} also reconstructs an explicit physical Hamiltonian and a four-dimensional generalized-extended-mimetic covariant completion for a representative anti-de Sitter-core model. The present paper uses the broader fixed-$\tilde f$ static-vacuum structure, rather than that specific anti-de Sitter-core realization, to construct inner-extremal regular black holes and analyze their behavior under colliding null shells.

In this work, we focus on the physically relevant case of $D=4$. A recent study in Ref.~\cite{FZ:26}, employing colliding null shells in the vicinity of the inner horizon, indicated a potential suppression of mass inflation. This result lends support to the viability of RBHs -- including non-inner extremal configurations -- as self-consistent solutions within modified gravity theories. Although the analysis presented there is largely model-independent and encompasses the case $D=4$, the explicit constructions were performed within the framework of quasitopological gravity. 

In the particular higher-dimensional polynomial quasitopological framework used for the explicit constructions of Ref.~\cite{FZ:26}, genuinely vacuum regular black-hole solutions are not obtained as four-dimensional polynomial quasitopological solutions \cite{BCH:19,BCHLM:22,MM:23,BHM:25}. Although other four-dimensional effective or nonpolynomial gravitational frameworks may realize vacuum regular black holes, this motivates studying the same mass-inflation question within an intrinsically four-dimensional polymerized-vacuum framework.

Here, we revisit the problem within a generic polymerized vacuum framework derived from an LTB collapse description. Our analysis encompasses both degenerate and non-degenerate inner horizons. Although the existence and detailed geometry of a given regular black hole depend on the chosen function $\tilde f$, the null-shell crossing analysis can be formulated across this class of models. A central assumption is that each of the four regions separated by the ingoing and outgoing null shells is described by a static, spherically symmetric vacuum solution of one and the same fixed polymerized theory: the function $\tilde f$ is identical in all four regions, while the mass parameter may differ from one region to another. In using the standard DHBI crossing relation as an effective matching condition for the metric sector, we further assume that the auxiliary reference-clock sector introduces no leading localized contribution at the shell-crossing sphere. Our conclusions should therefore be understood within this static four-region, metric-sector approximation. Within this restricted scope, the present analysis provides a useful setting in which to compare with, and test the extent to which, the behavior found in Ref.~\cite{FZ:26} persists in the polymerized vacuum framework. To implement this program, we first construct regular black hole solutions that admit both types of inner horizons. Although degenerate configurations have previously been obtained in quasitopological gravity \cite{FKK:25}, we explicitly demonstrate how such solutions can be realized in four-dimensional polymerized vacuum.

The remainder of this paper is organized as follows. In Sec.~\ref{sec:polymerized-vacuum}, we briefly review how RBHs can be derived as polymerized vacuum solutions. In Sec.~\ref{sec:non-degenerate:sol}, we present illustrative examples, focusing in particular on the Hayward model in the case of a non-degenerate inner horizon. In Sec.~\ref{sec:degenerate:sol}, we construct a specific inner-extremal RBH configuration and, in Sec.~\ref{sec:mass-tuning}, we demonstrate that such a solution can exist only under a finely-tuned mass parameter. In Sec.~\ref{sec:DHBI}, we review the DHBI method used to describe mass inflation in a general framework and derive a generic criterion establishing the conditions for mass inflation within the simplified colliding null-shell model. We then analyze this mechanism in the polymerized vacuum setting in Sec.~\ref{sec:mass-inflation}, applying our results to the two cases introduced earlier in Secs.~\ref{sec:DBHI:non-deg} and \ref{sec:DHBI:deg}. Finally, in Sec.~\ref{sec:discussion}, we summarize our main findings and outline possible directions for future research. Throughout this work, we restrict ourselves to four spacetime dimensions and adopt dimensionless units such that $\hbar = c = G = 1$.

\section{Polymerized vacuum solutions} \label{sec:polymerized-vacuum}

The polymerized vacuum solutions considered in this work originate from the LTB description of spherically symmetric gravitational collapse of pressureless dust \cite{LLB:07}. In this framework the spacetime is foliated by hypersurfaces comoving with the dust fluid, whose worldlines define a natural physical clock. In the reduced Hamiltonian description this dust time is used to deparameterize the dynamics. The vacuum branch used below is obtained by setting the dust density to zero, leaving a conserved mass parameter that labels the static solutions.

In these coordinates the dynamics of the system simplifies considerably: the evolution of the geometry can be described as a collection of independent spherical shells labeled by $x$, each characterized by two conserved quantities -- the Misner–Sharp mass $M(x)$ and the LTB energy function $\mathcal{E}(x)$, which represents the specific binding energy of the shell. The resulting equations of motion take the form of a generalized Friedmann equation for each shell, implying that the full inhomogeneous spacetime can be viewed as a continuum of homogeneous mini-superspace systems evolving independently. This decoupled structure can be made explicit in the Hamiltonian formulation, where the dynamics of each shell is generated by a shell Hamiltonian $H(b,v;\mathcal{E})$, written in terms of canonical variables related to the extrinsic curvature and the shell volume\footnote{The LTB construction begins with the metric line element
	\begin{align}
		ds^2=-dt^2+\frac{(E^{\phi})^2}{E^{x}}dx^2+E^{x}d\Omega^2.
	\end{align} 	
	Due to the decoupling of the dynamics for each shell labeled by $x$, it is convenient to introduce the canonical variables
	\begin{align}
		b(x)=\frac{K_{\phi}}{\sqrt{E^{x}}}, \quad v(x)=(E^{x})^{3/2},
	\end{align}
	that satisfy the Poisson bracket
	\begin{align}
		\left\{ b(x),v(y)\right\}=\frac{3}{2}\delta(x-y).
\end{align}
Here $K_{\phi}$ is the angular component of the extrinsic-curvature variable used in the spherically symmetric phase space. The curvature function $\mathcal{R}_1$ is given by $\mathcal{R}_1(t,x)\equiv -\tensor{R}{^{\theta\phi}_{\theta\phi}}$. A detailed discussion of this framework can be found in Refs.~\cite{LLB:07,GLSW:24,GLSW:25,GL:25,LS:26}.
}. 

Polymerized vacuum models arise when the classical shell Hamiltonian is modified in a way inspired by loop quantum gravity (LQG). The guiding principle is to preserve the structural features of the LTB framework -- namely the decoupling of the shell dynamics and spatial diffeomorphism invariance -- while allowing the functional form of the Hamiltonian to encode quantum-gravity corrections. Within this construction the function $H(b,v;\mathcal{E})$ is no longer restricted to its classical form but may contain nonlinear dependence on the curvature variables. Nevertheless, compatibility with the underlying geometry imposes strong constraints on its structure, ensuring that the modified theory still admits LTB-type solutions and can be lifted to a spatially covariant four-dimensional effective theory. In particular, these requirements imply that the Hamiltonian must factorize as $H(b,v,\mathcal{E})=vH_{0}(b,\mathcal{R}_1)$, where $\mathcal{R}_1$ represents the intrinsic curvature of the spatial slices. 

To obtain static black hole geometries, one considers the vacuum limit of the collapse model, corresponding to regions where the dust density vanishes and the Misner–Sharp mass becomes constant. In this case the spacetime admits a vacuum solution uniquely determined by the mass, resembling the role of Birkhoff’s theorem in general relativity. This is achieved by imposing a geometric condition that eliminates the residual dependence of the spacetime geometry on the LTB energy function $\mathcal{E}(x)$ and constrains the allowed form of the shell dynamics. As shown in Refs.~\cite{GL:25,LS:26}, these conditions uniquely determine the structure of the resulting static metric. In Schwarzschild-like coordinates the line element takes the form
\begin{align}
	ds^2=-f(r)dt^2+f(r)^{-1}dr^2+r^2d\Omega^2,\label{eq:ds}
\end{align}
with the metric function necessarily obeying
\begin{align}
	f(r)=1-r^2\tilde{f}\left(\frac{2m}{r^3}\right).\label{eq:f}
\end{align}
Here $m$ denotes the conserved integration constant of the static vacuum branch.\footnote{The word ``mass'' is used in two related but distinct senses in this work. The parameter $m$ is the conserved integration constant of the polymerized vacuum solution, equivalently the quantity entering $H_0=2m/r^3$ on the static branch of the reduced dynamics \cite{LS:26}. Independently, any spherically symmetric metric defines a geometric Misner--Sharp mass function,
$M_{\rm MS}^{\rm geom}(r)=r(1-f)/2= r^3\tilde f(2m/r^3)/2$. In the DHBI analysis below, mass inflation is diagnosed operationally by the growth of this geometric mass function, or equivalently by the growth of the metric function generated by the shell crossing. In the polymerized theory this should be regarded as a geometric instability diagnostic, not automatically as a blow-up of the conserved Hamiltonian mass or of a fundamental matter energy.}

In the vacuum branch considered here, setting the dust density to zero does not invalidate the dust-time deparameterized description. Rather, the dust is taken in a test/reference limit: it provides the clock and associated foliation but does not backreact on the geometry, while the mass parameter of the polymerized vacuum solution remains a conserved integration constant. Consequently, if the reference dust congruence develops caustics or the associated foliation breaks down, this should be interpreted as a failure of the chosen reference system, not by itself as a physical matter singularity or curvature singularity. Physical instabilities should instead be assessed through invariant curvature quantities, or through the behavior of the geometric and canonical variables entering the polymerized Hamiltonian. In the DHBI analysis below, the Misner--Sharp mass is therefore used in this operational geometric sense, as a diagnostic of local metric amplification rather than as the conserved Hamiltonian mass of the polymerized vacuum solution.

This expression represents the most general form of a static, unique vacuum solution within the polymerized LTB framework. The function $\tilde{f}(\mathcal{X})$, with $\mathcal{X}=2m/r^3$, encodes the effective modifications to the classical Schwarzschild solution. In the classical limit one recovers $\tilde{f}(\mathcal{X})=\mathcal{X}$, reproducing 
\begin{align}
	f_{\mathrm{S}}(r)=1-\frac{2m}{r},
\end{align}
while different choices of $\tilde{f}$ describe distinct effective models within the polymerized vacuum framework. 

The uniqueness statement should be understood within a fixed polymerized model. That is, once the function $\tilde f$ and the vacuum branch of the reference-dust construction have been specified, the static vacuum solution is uniquely characterized by the integration constant $m$.  Different choices of $\tilde f$ correspond to different effective polymerized theories, rather than to distinct solutions within one and the same theory.  In this sense, Eq.~\eqref{eq:f} gives the general form of the static vacuum solution within this class of models, while the choice of $\tilde f$ selects the particular effective theory under consideration.

Finally, we note that the Hamiltonian framework can be translated into a Lagrangian formulation \cite{GL:25,LS:26}. The resulting action belongs to the class of generalized extended mimetic gravity models and provides a manifestly covariant description of the preferred-foliation dynamics \cite{TK:17,LMNV:19}.  In this formulation the scalar field plays the role of the reference clock inherited from the dust-time deparameterization.  On the vacuum branch considered in this work the dust density is set to zero, so this scalar should not be interpreted as an independent propagating matter degree of freedom of the black-hole solution.  Rather, it enforces the preferred foliation/clock constraint used to covariantize the spatial Hamiltonian. Since the underlying
Hamiltonian evolution is second order in the preferred time, the corresponding covariant formulation contains no Ostrogradsky-type higher-time-derivative instability on this branch.

Having established the general structure of the vacuum solutions, we can now use Eq.~\eqref{eq:f} to construct explicit four-dimensional regular black hole geometries exhibiting both non-degenerate and degenerate horizons.

\subsection{Non-degenerate inner horizon} \label{sec:non-degenerate:sol}

In this section, we focus on RBHs that possess a non-degenerate inner horizon, namely, an inner horizon with non-vanishing surface gravity. Most well-known RBH models in the literature fall into this class. For the purposes of this paper, we adopt the Hayward model as a representative example to illustrate our arguments, although the methods developed here can be applied more generally to a broad range of RBH models. The Hayward model is characterized by the metric function
\begin{align}
	f_{\mathcal{H}}(r)=1-\frac{2mr^2}{r^3+2m\ell^2}.\label{eq:fh}
\end{align}
This model achieves regularization of the central region by effectively replacing the singular core with a de Sitter core. This is accomplished through the introduction of a minimal length scale $\ell$, which controls the deviation from the Schwarzschild geometry. In the limit $\ell \to 0$, the solution smoothly reduces to the singular Schwarzschild black hole. Such a model can be consistently generated within our framework derived from Eq.~\eqref{eq:f}. In particular, as shown in Refs.~\cite{GLSW:25,LS:26}, the Hayward black hole solution can be reproduced by appropriately choosing the function
\begin{align}
	\tilde{f}_{\mathcal{H}}(\mathcal{X})=\frac{\mathcal{X}}{1+\ell^2\mathcal{X}}.
\end{align}
This metric function admits two positive roots, $r_{+}$ and $r_{-}$, corresponding to the outer and inner horizons, respectively. For the purposes of this paper, we are interested in their large-mass expansions. One can show that, in the limit of large $m$, they take the form
\begin{align}
	r_{+}=2m-\frac{\ell^2}{2m}+\mathcal{O}(m^{-3}),\label{eq:rp-H}
\end{align}
while the inner horizon behaves as
\begin{align}
	r_{-}=\ell+\frac{\ell^2}{4m}+\mathcal{O}(m^{-2}).\label{eq:rm-H}
\end{align}
An expression that will be useful later is the surface gravity at the inner horizon in this large-mass limit. In general, the surface gravity is given by
\begin{align}
	\kappa_-=\frac{1}{2}\partial_{r}f_{\mathcal{H}}\bigg|_{r=r_{-}},
\end{align}
Evaluating this expression yields the following expansion:
\begin{align}
	\kappa_{-}=-\frac{1}{\ell}+\frac{1}{m}+\mathcal{O}(m^{-2}).\label{eq:kappa-H}
\end{align}
This confirms that the inner horizon is non-degenerate, as it is characterized by a non-vanishing surface gravity, $\kappa_{-} \neq 0$. It is worth mentioning that a Hayward-type solution with an anti-de Sitter core can also be constructed, and such configurations have been discussed in the literature \cite{ALNV:25}, including within the polymerized vacuum framework \cite{LS:26}. However, the resulting expressions are considerably more involved than in the de Sitter core case. For this reason, we restrict our analysis to the latter as a representative and technically transparent example. We now turn to the case of an RBH with a degenerate inner horizon.

\subsection{Degenerate inner horizon}\label{sec:degenerate:sol}

Having analyzed the non-degenerate case, we now turn to regular black hole geometries that admit degenerate inner horizons, commonly referred to as inner-extremal configurations. A general classification of RBH metrics with two horizons and their possible degeneracies can be found in Ref.~\cite{MS:23}. Following the definitions introduced there, we focus on the case of a triple-degenerate horizon as a representative example, since it constitutes the simplest non-trivial scenario beyond the non-degenerate case. This structure can be demonstrated by analyzing the behavior of the expansion of outgoing null rays in the vicinity of the horizon. Although horizons with higher-order degeneracy could, in principle, be accommodated within our framework, the corresponding analysis becomes increasingly cumbersome while providing limited additional physical insight. For this reason, we restrict our discussion to the triple-degenerate configuration.

To illustrate this case, we note that for an RBH with an outer horizon $r_{+}$ and a degenerate inner horizon $r_{-}$, the metric function must satisfy the following conditions. For the outer horizon,
\begin{align}
	f(r_{+})=0,
\end{align} 
while for the inner horizon we require
\begin{align}
	f(r_{-})=0,\quad f'(r_{-})=0, \quad f''(r_{-})=0.
\end{align}
The last two conditions characterize the degeneracy of the inner horizon. In particular, the vanishing of $f'(r_{-})$ implies a zero surface gravity, rendering the RBH inner-extremal, while the additional condition on the second derivative ensures the higher-order degeneracy we aim to model. 

Our goal is to construct a function $\tilde{f}$ that satisfies these requirements and thereby generates inner-extremal RBHs in $D=4$ within the polymerized vacuum framework. As a starting point, we assume a rational ansatz for $\tilde{f}$ in terms of polynomials. This choice provides sufficient flexibility to impose the necessary asymptotic conditions and regularity at the core. Specifically, we consider
\begin{align}
	\tilde{f}(\mathcal{X})=\frac{a_{0}+a_{1}\mathcal{X}+a_{2}\mathcal{X}^2}{b_{0}+b_{1}\mathcal{X}+b_{2}\mathcal{X}^2},
\end{align}
where we recall that $\mathcal{X} = 2m/r^{3}$, which is the only scalar through which modifications enter via the function $\tilde{f}$.

Using this ansatz, the large-$r$ expansion of the metric function yields
\begin{align}
	f(r)=-\frac{a_0}{b_0}r^2+1-\frac{2(a_1b_0-a_0b_1)m}{b^2_0r}+\mathcal{O}(r^{-4}).
\end{align}
Requiring Schwarzschild asymptotics immediately imposes $a_0 = 0$, which in turn implies $a_1 = b_0$. Without loss of generality, we fix the normalization by setting $a_1 = b_0 = 1$. These conditions are sufficient to guarantee the correct asymptotic behavior. With these constraints, the function $\tilde{f}$ reduces to
\begin{align}
	\tilde{f}(\mathcal{X})=\frac{\mathcal{X}(1+a_2 \mathcal{X})}{1+b_1 \mathcal{X}+b_2 \mathcal{X}^2}. \label{eq:ft-deg}
\end{align}
We next examine the behavior near the center to determine whether the geometry admits a regular core. Expanding for small $r$, we obtain
\begin{align}
	f(r)=1-\frac{a_2}{b_2}r^2+\mathcal{O}(r^5).
\end{align}
Thus, provided $b_2 \neq 0$ and $a_2 \neq 0$, the metric develops either a de Sitter or anti-de Sitter core, depending on the sign of $a_2/b_2$. In either case, the geometry remains regular at the center. Collecting these results, the metric function takes the form
\begin{align}
	f(r)=1-\frac{2m r^2(r^3+2a_2m)}{r^6+2b_1 m r^3+4 b_2 m^2}.\label{eq:f-deg-initial}
\end{align}
At this stage, the solution depends on four parameters.  We now restrict attention to the sector of parameter space in which the metric function has four positive real roots, denoted by $r_i$ with $i=1,2,3,4$, and in which the denominator of Eq.~\eqref{eq:f-deg-initial} has no positive real roots. The horizon radii satisfy $f(r_i)=0$, and these four algebraic equations can be used to trade the parameters $a_2$, $b_1$, $b_2$, and $m$ for the roots $r_i$. The resulting expressions are rather lengthy and are therefore presented in Appendix~\ref{sec:app}.  Thus, at this point we have selected the regular black-hole sector relevant for the subsequent construction, but we have not yet imposed any degeneracy condition.  We now proceed to specialize this sector to the case of a degenerate inner horizon.

\begin{figure}[!htbp]
	\centering
	\includegraphics[scale=0.45]{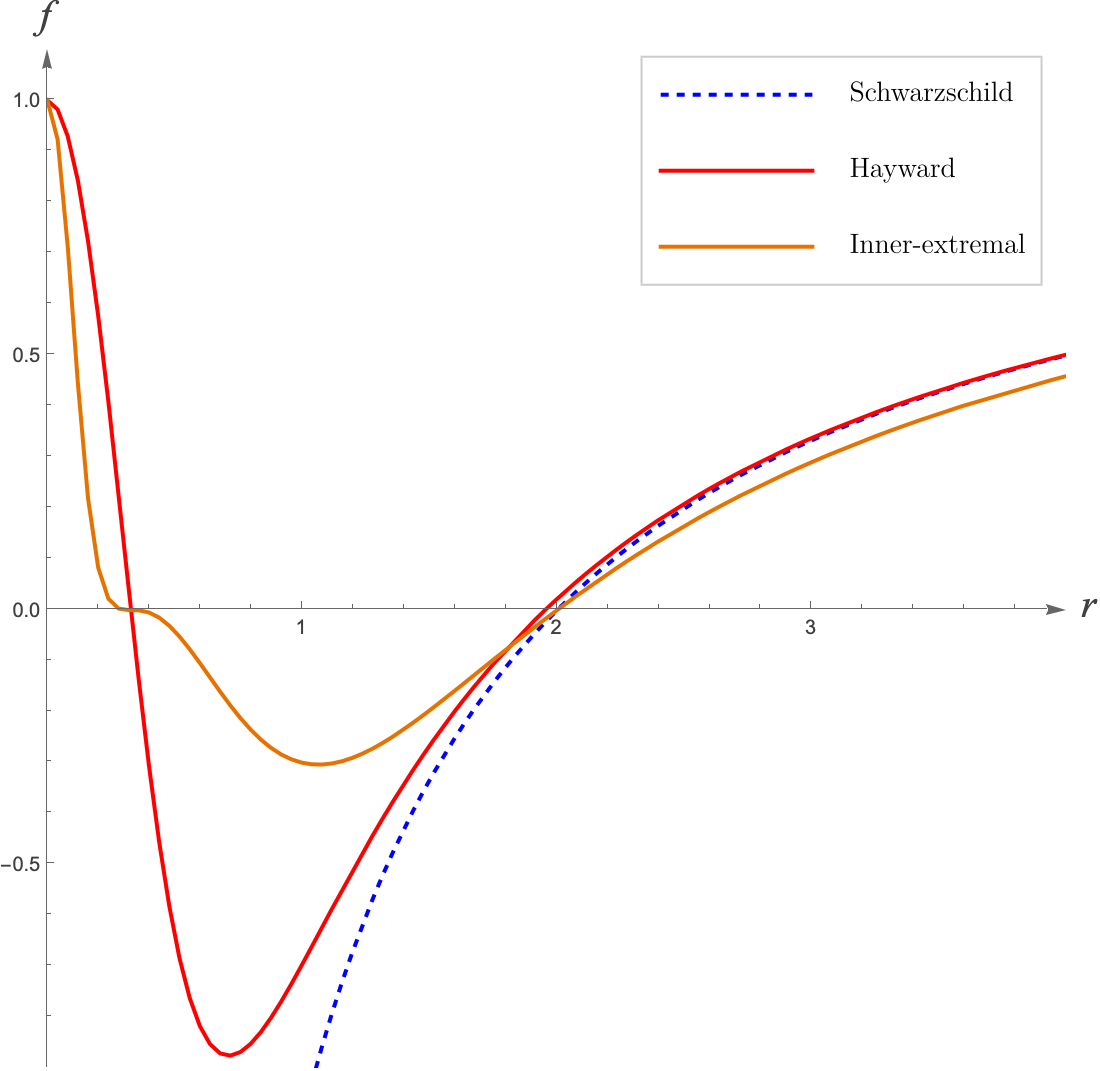}
	\caption{\textbf{Metric functions.} The metric functions for the singular Schwarzschild black hole and the RBHs of Hayward and the inner-extremal model are shown as functions of the areal radius $r$. For the Schwarzschild and Hayward cases, the mass parameter is fixed to $m=1$ and the minimal length scale to $\ell=0.3$. For a meaningful comparison, we choose $r_{+}=2$ in the inner-extremal model, while $r_{-}\simeq 0.328$ is selected to match the inner horizon of the Hayward solution for the same parameter values.}\label{fig:plotf}
\end{figure}

We now impose the condition of triple degeneracy at the inner horizon by identifying $r_{-}=r_{1}=r_{2}=r_{3}$ while the remaining root corresponds to the outer horizon, $r_{+}=r_{4}$. Fig.~\ref{fig:plotf} compares the metric functions of the Schwarzschild, Hayward, and inner-extremal black holes. Under this identification, the previously obtained relations simplify considerably, as the system is now fully characterized by the two independent scales $r_{-}$ and $r_{+}$. The expressions are given by:
\begin{align}
	a_2=\frac{r_{-}^3 \, (r_{-}^2 + 3 r_{-} r_{+} + r_{+}^2) \, (r_{-}^3 + 3 r_{-}^2 r_{+} + 6 r_{-} r_{+}^2 + 5 r_{+}^3)}{\bigl(2 r_{-}^3 + 6 r_{-}^2 r_{+} + 3 r_{-} r_{+}^2 + r_{+}^3\bigr)^2},\label{eq:a2s}
\end{align}
\begin{align}
	b_1=r_{-}^2 \left( 5 - \frac{8 r_{-} (r_{-}^2 + 3 r_{-} r_{+} + r_{+}^2)}{2 r_{-}^3 + 6 r_{-}^2 r_{+} + 3 r_{-} r_{+}^2 + r_{+}^3} \right),
\end{align}
\begin{align}
	b_2=\frac{r_{-}^5 \, r_{+}^2 \, (r_{-} + r_{+}) \, (r_{-}^2 + 3 r_{-} r_{+} + r_{+}^2)}{\bigl(2 r_{-}^3 + 6 r_{-}^2 r_{+} + 3 r_{-} r_{+}^2 + r_{+}^3\bigr)^2},
\end{align}
\begin{align}
	m=\frac{1}{2} \left( r_{+} + \frac{r_{-}^2 (2 r_{-} + 5 r_{+})}{r_{-}^2 + 3 r_{-} r_{+} + r_{+}^2} \right).
\end{align}

Imposing triple degeneracy restricts the allowed parameter space of the solution. In the parametrization used here, this restriction can be implemented by expressing the horizon radii $r_{+}$ and $r_{-}$ in terms of $b_1$ and $b_2$. Substituting these relations into Eq.~\eqref{eq:a2s} then fixes $a_2$ in terms of the same parameters. Thus the triple-degenerate configurations form a two-parameter family, which may be labelled by $b_1$ and $b_2$. An important implication is that the mass is no longer freely variable: once the inner-extremal condition is imposed, the mass is tuned to a specific value determined by these parameters.\footnote{Equivalently, the triple-degeneracy conditions define a two-dimensional submanifold in the parameter space $(a_2,b_1,b_2,m)$. One may use $(r_-,r_+)$ as auxiliary parameters on this submanifold, or, where the map can be inverted, use $(b_1,b_2)$ instead. In the latter description, fixing the theory parameters $b_1$ and $b_2$ determines both the required value of $a_2$ and the corresponding inner-extremal mass $m=m_{\rm ext}(b_1,b_2)$.} Although this feature resembles what occurs in quasitopological gravity -- where the metric function follows from an algebraic equation of motion -- the present result arises within a four-dimensional polymerized vacuum framework. The restriction on the mass therefore arises from the degeneracy conditions, together with the specific way these are implemented through the effective quantum-gravity corrections encoded in $\tilde{f}$.

As we demonstrate in the following subsection, this feature is not an artifact of the specific example considered here. Rather, for admissible analytic choices of $\tilde f$ with nonpathological isolated degeneracy roots, it is a generic property of inner-extremal RBHs within this framework. The parameters $b_1$ and $b_2$ therefore determine the admissible mass values compatible with a degenerate inner horizon, reflecting a structural constraint intrinsic to the theory. For comparison, in quasitopological gravity the metric function is obtained from an algebraic relation of the form
\begin{align}
	h(\psi)=\frac{2m}{r^{D-1}}, \quad D\geqslant 5,
\end{align}
with $\psi = (1 - f(r))/r^2$. The function $h(\psi)$ is constructed to possess suitable properties and typically arises from an infinite tower of higher-curvature corrections, encoding effective quantum-gravity effects \cite{BCH:25}. In our construction, the metric function is derived from Eq.~\eqref{eq:f} in a structurally similar manner. Schematically, $\tilde{f}$ plays a role analogous to $h^{-1}(m/r^{D-1})$, modulo dimensional differences. However, due to the dimensional constraints inherent in quasitopological gravity, vacuum inner-extremal RBHs cannot be realized in $D=4$, which is the physically relevant case. 

By contrast, within our framework — based on LTB framework and the polymerized vacuum — we obtain a fully covariant construction in four dimensions. This provides an explicit realization of a vacuum inner-extremal regular black hole in four dimensions within the polymerized-vacuum framework. Unlike the quasitopological constructions, which rely on higher-dimensional higher-curvature dynamics, the present solution arises directly from the
four-dimensional polymerized LTB vacuum ansatz of Eq.~\eqref{eq:f}. The covariant
Lagrangian embedding of the underlying class of models is taken from Refs.~\cite{GL:25,LS:26}; the new ingredient here is the construction of the inner-extremal black-hole solution and its subsequent null-shell analysis.

\subsubsection{Fine-tuning of the mass parameter}\label{sec:mass-tuning}

The proof of the finely tuned mass follows closely the strategy of Ref.~\cite{FKK:25}, owing to the structural similarity between the metric function in quasitopological gravity and the one obtained in our formulation. We assume the existence of a degenerate inner horizon located at some radius $r_{*}$. For convenience, we introduce the quantity $\mathcal{X}_{*}=2m/r^3_{*}$ and thus 
\begin{align}
	f(r_*)=1-r^2_{*}\tilde{f}(\mathcal{X}_*)=0,\label{eq:fminus}
\end{align}
\begin{align}
	f'(r_*)=\tilde{f}(\mathcal{X}_*)-\frac{3}{2}\mathcal{X}_* \tilde{f}'(\mathcal{X}_*)=0.\label{eq:fprime_minus}
\end{align}
If Eq.~\eqref{eq:fprime_minus} is interpreted as a differential equation with respect to $\mathcal{X}_{*}$ and formally integrated, one finds
\begin{align}
	\tilde{f}(\mathcal{X}_{*}) = C \mathcal{X}_{*}^{2/3},
\end{align}
where $C$ is an integration constant. However, this form is non-analytic at infinity, i.e., $\mathcal{X}=0$ and is therefore incompatible with the regularity requirements needed for a consistent polymerized vacuum construction \cite{GLSW:25,GLRSW:25,GL:25,LS:26}. For this reason, we instead interpret Eq.~\eqref{eq:fprime_minus} as a relation that holds only at the specific point $\mathcal{X}_{*}$ corresponding to the degenerate horizon.
Using Eq.~\eqref{eq:fminus}, we obtain
\begin{align}
	\tilde{f}(\mathcal{X}_{*}) = \frac{1}{r_{*}^{2}}
	\quad \Rightarrow \quad
	r_{*} = [\tilde{f}(\mathcal{X}_{*})]^{-1/2}.
\end{align}
Combining this with the definition $\mathcal{X}_{*} = 2m/r_{*}^{3}$ yields
\begin{align}
	m = \frac{\mathcal{X}_{*}}{2} [\tilde{f}(\mathcal{X}_{*})]^{-3/2}.
\end{align}

Thus the inner-extremal condition does not leave a continuously
adjustable mass parameter within one fixed polymerized theory.  For a
fixed choice of $\tilde f$, the degenerate configuration occurs only for
special tuned values of the mass. This distinction will be important
below when we consider null shells, since a shell perturbation changes
the integration constant $m$ while the function $\tilde f$ is kept fixed.

\section{DHBI Formalism}\label{sec:DHBI}

In this section, we analyze the phenomenon of mass inflation within the framework of the Dray--'t Hooft--Barrabes--Israel (DHBI) formalism \cite{DH:85,BI:91}. In the present work, we adopt the operational viewpoint that mass inflation corresponds to an unbounded growth of the local Misner--Sharp mass in the vicinity of the Cauchy horizon. This should be understood as a measure of local geometric/metric amplification, not as a divergence of the conserved polymerized mass parameter $m$ that labels the vacuum solution. To probe this effect, we employ the standard colliding null-shell setup, consisting of one ingoing and one outgoing thin null shell intersecting near the inner horizon. As the metric function approaches zero at this surface, extreme blueshift effects arise. Radiation propagating along one null direction appears increasingly energetic to observers moving along the opposite direction. This mutual amplification generates a feedback mechanism whereby small perturbations can induce large gravitational backreaction. It is precisely this mechanism that drives mass inflation \cite{HA:10}.

\begin{figure}[t]
	\centering
	\begin{tikzpicture}[
		scale=2.3,
		line cap=round,
		line join=round,
		>=Latex
		]
		
		% ---- parameters
		\def\a{1.0}
		\def\L{0.8}
		
		% ---- outer diamond
		\draw[gray, line width=1.1pt] (-\a,0) -- (0,\a);
		\draw[gray, line width=1.1pt] (0,\a) -- (\a,0);
		\draw[black, line width=1.1pt] (\a,0) -- (0,-\a);
		\draw[black, line width=1.1pt] (0,-\a) -- (-\a,0);
		
		% ---- null rays
		\draw[blue, line width=1.2pt, ->] (\L,-\L) -- (-\L,\L);
		\draw[red, line width=1.2pt, ->] (-\L,-\L) -- (\L,\L);
		
		% ---- intersection
		\fill (0,0) circle (0.035);
		
		% ---- regions
		\node at (0, 0.6) {$\mA$};
		\node at (0,-0.6) {$\mB$};
		\node at (0.6,0) {$\mC$};
		\node at (-0.6,0) {$\mD$};
		
		% ---- collision label
		\node at (0.15,0) {$\mathrm{S}$};
		
		% ---- horizons
		\node[rotate=45, sloped, above, gray] at ($(-\a,0)!0.80!(0,\a)$) {$r=r_{-}$};
		\node[rotate=-45, sloped, above, gray] at ($( \a,0)!0.80!(0,\a)$) {$r=r_{-}$};
		
		\node[sloped, below, rotate=-45] at ($(-\a,0)!0.80!(0,-\a)$) {$r=r_{+}$};
		\node[sloped, below, rotate=45] at ($( \a,0)!0.80!(0,-\a)$) {$r=r_{+}$};
		
	\end{tikzpicture}
	
	\caption{\textbf{Null shells.} Penrose diagram illustrating the collision of an ingoing (blue) and an outgoing (red) null shell at the sphere $\mathrm{S}$. The spacetime is divided into four regions, $\mA$, $\mB$, $\mC$, and $\mD$. Region $\mB$ corresponds to the original background geometry, before the crossing of any shells. Region $\mC$ represents the geometry after the ingoing shell has crossed, while region $\mD$ corresponds to the geometry after the outgoing shell has crossed. The region of primary interest is $\mA$, which describes the spacetime after both shells have crossed, i.e., after their collision. The outer horizon $r=r_{+}$ forms the lower boundary of the interior region and is depicted as a solid black line, while the inner horizon $r=r_{-}$ forms the upper boundary and is illustrated with gray lines. Throughout this work, we assume that the sphere $\mathrm{S}$ lies close to the inner horizon $r_-$.}
	\label{fig:penrose}
\end{figure}
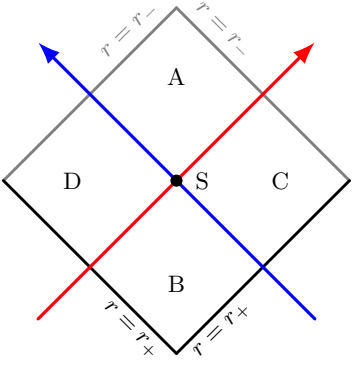

As emphasized in Sec.~\ref{sec:intro}, a  key assumption underlying our analysis is that each of the four spacetime regions separated by the null shells is described by a static, spherically symmetric geometry of the form given in Eq.~\eqref{eq:ds}, with metric function specified by Eq.~\eqref{eq:f} (see Fig.~\ref{fig:penrose}). While the functional structure of the metric is identical in all regions, the mass parameter differs from quadrant to quadrant. This assumption is justified by the fact that the geometry arises as a polymerized vacuum solution admitting a Birkhoff-type theorem, and is therefore uniquely characterized by a single integration constant, the mass $m$. All modifications to the classical Schwarzschild structure are encoded in the function $\tilde{f}$, which remains fixed across the spacetime. Since the regions separated by the shells are vacuum domains, it is natural to model them as distinct solutions of the same effective theory, differing only in their respective mass parameters.

The interaction of the shells is governed by the DHBI matching conditions. Although the shells carry distributional stress-energy, the spacetime metric itself remains continuous across each null hypersurface, while its derivatives are piecewise continuously differentiable. In coordinate systems that are continuous across the shells, the metric components $g_{\mu\nu}$ are continuous, ensuring a consistent geometric description of the crossing sphere. The essential dynamics emerges at the intersection point of the shells. The DHBI formalism provides an algebraic relation between the metric functions in the four regions. This relation determines whether the local Misner-Sharp diagnostic, or equivalently the metric function inferred at the crossing sphere, undergoes large amplification as the collision approaches the inner horizon. The explicit relation takes the form
\begin{align}
	f_{\mA}=\frac{f_{\mC}f_{\mD}}{f_{\mB}},\label{eq:DBHI}
\end{align}
where $f_\mA$, $f_\mB$, $f_\mC$, and $f_\mD$ denote the metric functions in the corresponding spacetime quadrants. 

This relation is a local matching condition at the crossing sphere and is largely kinematical: it follows from the continuity of the metric across the null shells and from the use of coordinates adapted to the four static spherically symmetric regions.  In particular, it does not rely on the Einstein equations in their standard form. For this reason it has also been used as the null-shell matching condition in modified gravity settings, including the quasitopological-gravity analysis of Ref.~\cite{FZ:26}. Following this approach, we use the DHBI relation as an effective thin-shell matching condition for the metric sector of the polymerized vacuum solutions considered here\footnote{In resummed quasitopological gravity, the regular vacuum black-hole solutions examined in Ref.~\cite{DKS:26} were found not to admit spherical null thin shells, illustrating the theory dependence of null-shell junction conditions.}. In a complete treatment of the covariantized theory, the reference-clock sector could in principle modify the junction conditions or introduce additional localized data at the shell-crossing sphere. We assume that such contributions are absent, or subleading for the near-horizon metric-sector diagnostic considered here. Under this assumption, the crossing relation takes the standard DHBI algebraic form.

\begin{table*}[t]
	\centering
	\renewcommand{\arraystretch}{1.8}
	\setlength{\tabcolsep}{12pt}
	\begin{tabular}{|c||c||c|}
		\hline
		\rule{0pt}{2.6ex} $2q-k$ 
		& $ f_\mA$ 
		& Mass Inflation \rule{0pt}{2.6ex} \\
		\hline\hline
		\rule{0pt}{2.6ex} $2q-k<0$ 
		& $\sim x^{\,2q-k} \rightarrow \infty$ as $x\to 0$ 
		& $\cmark$  \rule{0pt}{2.6ex} \\
		\hline
		\rule{0pt}{2.6ex} $2q-k=0$ 
		& finite non-vanishing contribution 
		& marginal case \rule{0pt}{2.6ex} \\
		\hline
		\rule{0pt}{2.6ex} $2q-k>0$ 
		& $\sim x^{\,2q-k} \rightarrow 0$ as $x\to 0$ 
		& $\xmark$ \rule{0pt}{2.6ex} \\
		\hline
	\end{tabular}
	\caption{\textbf{Mass inflation criterion.} Classification of the near-horizon behavior of the crossing term 
		$f_\mA\sim \delta_\mC\delta_\mD/f_\mB \sim x^{2q-k}$. 
		The sign of the exponent $2q-k$ determines whether mass inflation occurs. 
		A negative exponent leads to divergence at the inner horizon ($x\to 0$), 
		a positive exponent suppresses the effect, while the vanishing case corresponds to a marginal configuration.}
	\label{table:mass-inflation}
\end{table*}

In the following, we implement this construction in a dimensionless parametrization and establish a general criterion for mass inflation based on the near-horizon expansions of the metric function and its dependence on the mass parameter. We closely follow the strategy of Ref.~\cite{FZ:26}, adopting their notation and introducing dimensionless variables that considerably simplify the analysis. In particular, we define a dimensionless radial coordinate $\rho$ and a parameter $\beta$, which allow us to probe the geometry near the inner horizon through $\rho$ while keeping explicit control of the mass dependence of $f$ through $\beta$. These quantities are defined as
\begin{align}
	\rho=\frac{r}{r_0}, \quad \text{and} \quad \beta=\frac{r_0}{2m},
\end{align}
where the scale $r_0$ is defined as the large-mass limit of the inner-horizon radius, namely
\begin{align}
	\lim_{m\to \infty}r_{-}=r_0.\label{eq:r0}
\end{align}
We now turn to the description of the null shells. Let us assume that the ingoing and outgoing shells carry masses $m_{\mathrm{in}}$ and $m_{\mathrm{out}}$, respectively. It is convenient to introduce the corresponding dimensionless parameters defined by
\begin{align}
\mu_\mC=\frac{m_{\mathrm{in}}}{m+m_{\mathrm{in}}}, \quad \mu_\mD=\frac{m_{\mathrm{out}}}{m+m_{\mathrm{out}}},
\end{align}
which quantify the relative strength of the perturbations with respect to the background mass $m$. With these definitions, the mass parameters in regions $\mC$ and $\mD$ can be written as
\begin{align}
	m_\mC=\frac{m}{1-\mu_{\mC}}, \quad m_\mD=\frac{m}{1-\mu_\mD}.
\end{align}
It follows directly that the corresponding dimensionless parameters become
\begin{align}
	\beta_\mC=\beta(1-\mu_\mC), \quad \beta_\mD=\beta(1-\mu_\mD).
\end{align}
Using these relations, the metric functions in the different regions can be expanded around the background configuration $f(\rho,\beta)$ as
\begin{align}
	f_{\mC}\equiv f(\rho,\beta_\mC)\simeq f(\rho,\beta)-\beta\mu_\mC\partial_\beta f+\cdots, \label{eq:fc-initial}
\end{align}
and 
\begin{align}
	f_{\mD}\equiv f(\rho,\beta_\mD)\simeq f(\rho,\beta)-\beta\mu_\mD\partial_\beta f+\cdots. \label{eq:fd-initial}
\end{align}
Using the crossing-sphere relation given in Eq.~\eqref{eq:DBHI}, we introduce the near–inner-horizon variable $x=\rho-\rho_{-}$, where $\rho_{-}$ denotes the dimensionless location of the inner horizon. Expanding the metric function in region $\mB$ around $\rho=\rho_{-}$ yields
\begin{align}
	f_{\mB}(\rho,\beta)\simeq f_{\mB}(\rho_{-}(\beta),\beta)+\sigma_{k}x^k+\cdots, \quad \sigma_{k}\neq 0.
\end{align}
Since $\rho_{-}$ corresponds to the inner horizon, the leading term vanishes, and therefore
\begin{align}
		f_{\mB}(\rho,\beta)\simeq \sigma_{k}x^k+\cdots, \quad \sigma_{k}\neq 0,\label{eq:fB}
\end{align} 
where $k>0$ encodes the order of degeneracy of the inner horizon. In addition, we must quantify how sensitively the metric function responds to variations of the mass parameter, or equivalently to $\beta$. This is captured by expanding the derivative of $f$ with respect to $\beta$ near the inner horizon:
\begin{align}
	\partial_{\beta}f(\rho,\beta)=\tau_{q}(\beta)x^{q}+\cdots, \quad \tau_{q}(\beta)\neq 0. \label{eq:partial-f-beta}
\end{align}
Here, $q\geqslant 0$ characterizes the leading-order dependence of the $\beta$-variation on the distance from the horizon. Combining these near-horizon expansions allows us to determine the behavior of the crossing-sphere relation and to identify the potentially divergent contributions. More specifically, the expansions in Eqs.~\eqref{eq:fc-initial} and \eqref{eq:fd-initial}, together with the identification $f(\rho,\beta)=f_\mB$, can be rewritten near the inner horizon as
\begin{align}
	f_\mC=f_\mB-\beta \mu_\mC\partial_{\beta}f\simeq \sigma_{k}x^k-\beta \mu_\mC \tau_qx^q,
\end{align}
\begin{align}
	f_\mD=f_\mB-\beta \mu_\mD\partial_{\beta}f\simeq \sigma_{k}x^k-\beta\mu_\mD \tau_qx^q,
\end{align}
where $\sigma_k$ and $\tau_q$ are the leading coefficients of the respective expansions. For convenience, we introduce the deviations
\begin{align}
	\delta_{\mC}=f_\mC-f_\mB\simeq -\beta \mu_\mC \tau_qx^q,\label{eq:delta-c}
\end{align}
\begin{align}
	\delta_{\mD}=f_\mD-f_\mB\simeq -\beta \mu_\mD \tau_qx^q,\label{eq:delta-d}
\end{align}
which measure the departure from the background configuration in region $\mB$. Substituting Eqs.~\eqref{eq:delta-c} and \eqref{eq:delta-d} into the crossing relation \eqref{eq:DBHI}, we obtain
\begin{align}
	f_\mA=f_\mB+\delta_{\mC}+\delta_{\mD}+\frac{\delta_\mC\delta_{\mD}}{f_\mB}. \label{eq:fA}
\end{align}
After the collision of the two shells, the geometry in region $\mA$ is therefore described by $f_\mA$. From Eq.~\eqref{eq:fA}, it is clear that the only potentially divergent contribution arises from the last term. Using the near-horizon expansions of Eqs.~\eqref{eq:delta-c}, \eqref{eq:delta-d}, and Eq.~\eqref{eq:fB}, we find
\begin{align}
\frac{\delta_\mC\delta_{\mD}}{f_\mB}=\beta^2\mu_\mC\mu_\mD\frac{\tau^2_q}{\sigma_k}x^{2q-k}. \label{eq:DHBI-div}
\end{align}
Thus, even within this simplified colliding-shell model, the presence or absence of mass inflation near the inner horizon is determined by the exponent of $x$. If $2q-k<0$, the term diverges as $x\to 0$, signaling mass inflation. If $2q-k>0$, the contribution vanishes and no classical mass inflation occurs. The marginal case $2q-k=0$ corresponds to a finite but nonvanishing correction. These possibilities are summarized in Table~\ref{table:mass-inflation}. Up to this stage, the analysis has remained completely general. We now specialize to the polymerized vacuum framework and investigate the possibility of mass inflation for the specific RBH solutions constructed in Sec.~\ref{sec:polymerized-vacuum}.

\subsection{Mass inflation in polymerized vacuum}\label{sec:mass-inflation}

As discussed previously, the polymerized vacuum solutions provide a suitable setting for the application of the DHBI formalism developed in the preceding section. We now implement this framework explicitly and derive the expression for the potentially divergent contribution appearing in Eq.~\eqref{eq:fA}. We first consider an RBH with a non-degenerate inner horizon, using the Hayward model constructed in Sec.~\ref{sec:non-degenerate:sol} as our prototype. We then analyze the inner-extremal case, concentrating on the configuration obtained as a polymerized vacuum solution in Sec.~\ref{sec:degenerate:sol}.

\subsubsection{Non-degenerate inner horizon}\label{sec:DBHI:non-deg}

For the Hayward model described by Eq.~\eqref{eq:fh}, one observes that the parameter $r_0$ coincides with the minimal length scale $\ell$. This follows by taking the large-mass limit of the inner-horizon radius given in Eq.~\eqref{eq:rm-H}, which precisely yields the definition of $r_0$ introduced in Eq.~\eqref{eq:r0}. In this model, therefore, we have $r_0=\ell$. We now proceed by expressing the relevant quantities in terms of the dimensionless parameters
\begin{align}
	\rho=\frac{r}{r_0}=\frac{r}{\ell}, \quad \beta=\frac{r_0}{2m}=\frac{\ell}{2m}.
\end{align}
In these variables, the Hayward metric function takes the form
\begin{align}
	f_{\mathcal{H}}(\rho,\beta)=1-\frac{\rho^2}{1+\beta \rho^3}.
\end{align} 
We begin with the near-horizon expansion \eqref{eq:fB}. For the Hayward model one finds that the relevant expansion parameters are
\begin{align}
	k=1, \quad \sigma_{1}=-2+4\beta+\mathcal{O}(\beta^2).
\end{align}
As can be seen directly from Eq.~\eqref{eq:kappa-H}, the coefficient $\sigma_1$ corresponds to $2r_0\kappa_{-}$ expressed in dimensionless variables. Next, we analyze the $\beta$-dependence through the expansion \eqref{eq:partial-f-beta}, which yields
\begin{align}
	q=0, \quad \tau_{0}=1+\frac{\beta}{2}+\mathcal{O}(\beta^2). 
\end{align}
Combining these results, the potentially divergent contribution in Eq.~\eqref{eq:fA}, in the large-mass limit and for shell intersection sufficiently close to the inner horizon, becomes
\begin{align}
	\frac{\delta_\mC\delta_{\mD}}{f_\mB}\simeq -\frac{1}{2}\beta^2\mu_\mC\mu_\mD  \frac{1}{x}.
\end{align}
This expression diverges as $x\to 0$, giving the standard DHBI mass-inflation signal associated with non-degenerate inner horizons, as in the Hayward model. Here this signal is tracked by the local geometric Misner--Sharp mass, or equivalently by the metric function $f_\mA$ inferred at the crossing sphere, rather than by an unbounded growth of the conserved polymerized integration constant or by solving for an admissible asymptotic mass parameter in region $\mA$. As expected, the result agrees with those found in the literature and is equivalent to the expression derived in Ref.~\cite{FZ:26}. It is natural to ask at what distance from the inner horizon the colliding shells produce a significant amplification. As discussed in detail in Ref.~\cite{FZ:26}, this occurs when $f_\mA$ becomes of order $\mathcal{O}(1)$. This condition implies

\begin{align}
	x \sim \frac{1}{2}\mu_\mC\mu_\mD \beta^2,
\end{align}
which restoring dimensions, translates to
\begin{align}
	\Delta r\sim \frac{1}{2}\mu_\mC\mu_\mD \beta^2 \ell.
\end{align}
Here, $\Delta r$ denotes the areal-radius separation between the shell-intersection point and the inner horizon, rather than the corresponding local proper distance. In order to address this question within the polymerized vacuum framework, we must relate the minimal length scale $\ell$ to a fundamental quantity of the underlying quantum gravity description. A finite truncation of the Hayward model expressed as a geometric series has been derived in Refs.~\cite{GLSW:25,LS:26}. Within this approximation, one can establish a connection between the minimal length scale and the LQG area gap,

\begin{align}
\Delta=4\pi\sqrt{3}\gamma\ell^2_{p},
\end{align}
which arises from the spectrum of the LQG area operator \cite{Rov:10,RV:15,T:10}. This relation can also be inferred from an asymptotic expansion of the Hayward metric function. More concretely, a LQG–inspired effective model (see Refs.~\cite{GLSW:24,GLSW:25}) is characterized by a metric function of the form
\begin{align}
	f_{\mathrm{LQG}}(r)=1-\frac{2m}{r}+\frac{4m^2\alpha^2_\Delta}{r^4},
\end{align}
where the polymer parameter is defined as $\alpha_\Delta=\gamma\sqrt{\Delta}$. On the other hand, the large-$r$ asymptotic expansion of the Hayward metric yields
\begin{align}
	f_{\mathcal{H}}(r)=1-\frac{2m}{r}+\frac{4m^2\ell^2}{r^4}+\mathcal{O}(r^{-7}).
\end{align}
Comparing the two expressions suggests that the Hayward regularization parameter $\ell$ can be identified with the LQG-inspired scale $\alpha_\Delta$, so that $\ell=\alpha_\Delta$. This identification is heuristic: it follows from matching the leading large-$r$ correction and should therefore be understood as relating the effective Hayward regularization scale to the polymerization scale of the LQG-inspired model \cite{Giesel:2022rxi,GLSW:24,GLSW:25}. In this sense, the comparison provides a concrete link between the regularization scale appearing in the effective polymerized vacuum geometry and the fundamental LQG area gap. With this identification, the distance from the inner horizon at which mass inflation becomes significant is
\begin{align}
	\Delta r \simeq \left(\frac{3^{3/4}\pi^{3/2}\mu_\mC\mu_\mD\gamma^{9/2}\ell^2_{p}}{m^2}\right)\ell_{p}.
\end{align}

This result is significant for the following reason: If one assumes, as is customary in LQG, that the Barbero–Immirzi parameter $\gamma$ is of order $\mathcal{O}(1)$, then for macroscopic black holes ($m\gg \ell_p$) the dimensionless factor in the brackets is extremely small. Consequently,
\begin{align}
	\Delta r\ll \ell_{p}.
\end{align}
In other words, within the DHBI estimate, an order-one amplification of the Misner--Sharp mass would require the shell collision to occur at distances well below the Planck length. This regime lies outside the expected domain of validity of the classical or semiclassical geometric description employed here. Thus, in the same operational sense as Ref.~\cite{FZ:26}, the classical mass-inflation signal is strongly suppressed: the formal divergence is displaced to a trans-Planckian region where the thin-shell effective description cannot be trusted. The present result is therefore consistent with the quasitopological-gravity analysis of Ref.~\cite{FZ:26}, while differing in the origin of the relevant scale: in the polymerized vacuum model, the regularization parameter is tied to the LQG area gap, so the suppression scale is directly linked to the underlying quantum-geometric input.

\begin{figure}[!htbp]
	\centering
	\includegraphics[scale=0.45]{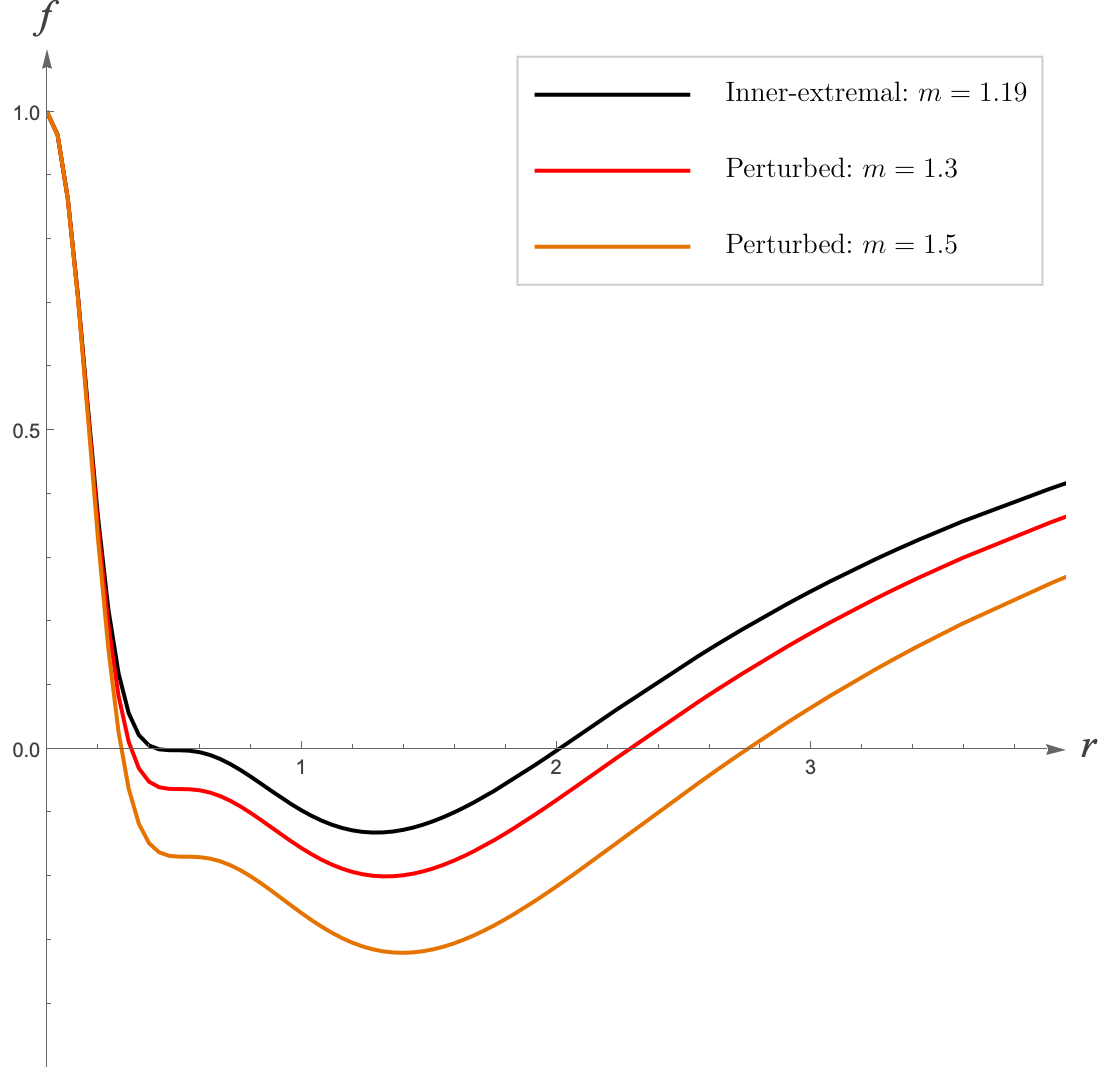}
\caption{\textbf{Inner-extremal mass perturbation.} Metric function $f(r)$ of Eq.~\eqref{eq:f-deg-initial} for a polymerized vacuum RBH theory with parameters $b_1 = 0.82971$ and $b_2 = 0.00761395$. The black curve corresponds to the finely tuned inner-extremal solution, with dependent parameter $a_2 = 0.163319$ and mass $m = 1.18966$; the horizons are located at $r_+ = 2$ and $r_- = 0.5$. The red and orange curves show small mass perturbations of the inner-extremal black hole for the same polymerized theory (i.e., with $a_2$, $b_1$, and $b_2$ fixed). These perturbations destroy the triple-degeneracy of the inner horizon, so the geometry is no longer inner-extremal.} 
\label{fig:plotfinext}
\end{figure}

\subsubsection{Degenerate inner horizon}\label{sec:DHBI:deg}
We now turn to the inner-extremal solution constructed in Sec.~\ref{sec:degenerate:sol}. This case has to be treated separately from the Hayward example. In the Hayward geometry the regularization scale is fixed and the mass parameter can be varied continuously, so the neighboring DHBI regions are naturally described by the same family of metrics with different
values of $m$. For the inner-extremal polymerized solution, however,
the degeneracy condition fixes the allowed mass once the function $\tilde{f}$ and its parameters have been chosen. Therefore, in the DHBI perturbation appropriate to a single polymerized theory, the null shells change only the integration constant $m$, while $\tilde{f}$ is kept fixed. The neighboring regions $\mC$ and $\mD$ then remain solutions of the same polymerized-vacuum theory, but they are no longer on the degenerate submanifold.  This is illustrated in Fig.~\ref{fig:plotfinext}: after a small mass perturbation at fixed theory parameters, the triple inner horizon of the tuned background is destroyed.

It is useful to state explicitly which metric function is being expanded.  Choose a tuned inner-extremal background with triple inner horizon $r_\ast$, outer horizon $r_{+\ast}$, and mass $m_\ast$. Let $a_{2\ast}$, $b_{1\ast}$, and $b_{2\ast}$ denote the values of the parameters in Eq.~\eqref{eq:f-deg-initial}, obtained after setting $r_-=r_\ast$ and $r_+=r_{+\ast}$. In the fixed-theory DHBI perturbation these three parameters are kept fixed, while the mass parameter is varied. The one-parameter metric function probed by the shells is therefore
\begin{align}
	f_{\rm fix}(r;m)=1-\frac{2m r^2\left(r^3+2a_{2\ast}m\right)}{r^6+2b_{1\ast}mr^3+4b_{2\ast}m^2}.
\end{align}
The background region $\mB$ is described by $f_\mB(r)=f_{\rm fix}(r;m_\ast)$, while the neighboring regions are obtained by replacing $m_\ast$ with the corresponding shell-shifted masses $m_\mC$ and $m_\mD$. This is the precise sense in which the variation is performed at fixed polymerized theory.

We now use the same dimensionless variables as above,
\begin{align}
	\rho=\frac{r}{r_0}, \quad \beta=\frac{r_0}{2m},
\end{align}
and denote the background value by $\beta_\ast=r_0/(2m_\ast)$. If
\begin{align}
A_\ast=\frac{a_{2\ast}}{r_0^2},\quad
B_{1\ast}=\frac{b_{1\ast}}{r_0^2},\quad
B_{2\ast}=\frac{b_{2\ast}}{r_0^4},
\end{align}
then the same fixed-theory metric can be written as
\begin{align}
	f_{\rm fix}(\rho,\beta)=1-\frac{\rho^2\left(\beta\rho^3+A_\ast\right)}{\beta^2\rho^6+\beta B_{1\ast}\rho^3+B_{2\ast}}.
\end{align}
This is just Eq.~\eqref{eq:f-deg-initial} rewritten in terms of $\rho$ and $\beta$,
with $A_\ast$, $B_{1\ast}$, and $B_{2\ast}$ held fixed. 

We write $\rho_\ast=r_\ast/r_0$ and expand around the tuned inner horizon using
\begin{align}
	x=\rho-\rho_\ast.
\end{align}
Since the background horizon is triple degenerate, the Taylor expansion
of $f_{\rm fix}(\rho,\beta_\ast)$ starts at cubic order:
\begin{align}
	f_\mB(\rho,\beta_\ast)=\sigma_3 x^3+O(x^4), \quad \sigma_3\neq 0.
\end{align}
Here $\sigma_3$ is not an additional assumption; it is the Taylor coefficient of the explicit metric function above,
\begin{align}
	\sigma_3=\frac{1}{6}\left.\partial_\rho^3 f_{\rm fix}(\rho,\beta_\ast)	\right|_{\rho=\rho_\ast}=\frac{r_0^3}{6}\left.\partial_r^3 f_{\rm fix}(r;m_\ast)\right|_{r=r_\ast}.
\end{align}
For the explicit large-mass background considered in Sec.~\ref{sec:degenerate:sol},
\begin{align}
	\rho_\ast=1+c\beta_\ast+O(\beta_\ast^2),\quad\sigma_3=-\frac{5}{6}+\frac{1}{4}(3+10c)\beta_\ast+\mathcal{O}(\beta_\ast^2).
\end{align}
The constant $c$ parametrizes the large-mass approach of the tuned inner-horizon radius to its limiting value. This part of the expansion depends only on the tuned background and is independent of how the neighboring DHBI quadrants are generated.

The relevant issue is the expansion of the mass variation. To make the fixed-theory derivative explicit, we define
\begin{align}
	F(Y)=r_0^2\,\tilde f\!\left(\frac{Y}{r_0^2}\right),
	\quad
	Y=\frac{1}{\beta\rho^3}.
\end{align}
The metric function can then be written as
\begin{align}
	f(\rho,\beta)=1-\rho^2 F(Y).
\end{align}
In the fixed-theory DHBI variation, $F$ is held fixed when differentiating with respect to $\beta$. At the tuned horizon the triple-degeneracy conditions are
\begin{align}
	f(\rho_\ast,\beta_\ast)=0,\quad
	\partial_\rho f(\rho_\ast,\beta_\ast)=0,\quad
	\partial_\rho^2 f(\rho_\ast,\beta_\ast)=0.
\end{align}
Equivalently,
\begin{align}
	F_\ast=\frac{1}{\rho_\ast^2},\quad 3Y_\ast F'_\ast=2F_\ast,
\end{align}
and
\begin{align}
	F'_\ast+3Y_\ast F''_\ast=0,
\end{align}
where primes denote derivatives with respect to $Y$, evaluated at
$Y=Y_\ast=1/(\beta_\ast\rho_\ast^3)$.

We now expand the fixed-theory derivative near the tuned horizon,
\begin{align}
	\partial_\beta f=\tau_0+\tau_1 x+\tau_2 x^2+\mathcal{O}(x^3).
\end{align}
Equivalently, the coefficients are
\begin{align}
	\tau_n=\frac{1}{n!}\left.\partial_\rho^n\partial_\beta
	f_{\rm fix}(\rho,\beta)\right|_{\rho=\rho_\ast,\beta=\beta_\ast},\quad n=0,1,2,\ldots.
\end{align}
In dimensional variables, the $\beta$-derivative at fixed $\rho$ is the mass derivative of the same fixed-theory function,
\begin{align}
	\partial_\beta f_{\rm fix}=-\frac{m}{\beta}\,\partial_m f_{\rm fix}(r;m),
	\quad r=r_0\rho,
\end{align}
with $a_{2\ast}$, $b_{1\ast}$, and $b_{2\ast}$ held fixed. At fixed $F$ and fixed $\rho$,
\begin{align}
	\partial_\beta f=\frac{1}{\beta^2\rho}F'(Y).
\end{align}
Taking one further $\rho$-derivative gives
\begin{align}
	\partial_\rho\partial_\beta f=-\frac{1}{\beta^2\rho^2}\left(F'(Y)+3YF''(Y)\right),
\end{align}
which vanishes at the tuned horizon by the degeneracy condition above.
The second derivative is
\begin{align}
	\left.\partial_\rho^2\partial_\beta f\right|_{\rho=\rho_\ast}=
	\frac{3}{\beta_\ast^2}
	\frac{Y_\ast}{\rho_\ast^3}
	\left(4F^{\prime\prime}_\ast
	+3Y_\ast F^{\prime\prime\prime}_\ast\right).
\end{align}
Therefore
\begin{align}
	\tau_0=\frac{1}{\beta_\ast^2\rho_\ast}F'_\ast
	=\frac{2}{3\beta_\ast},
	\quad
	\tau_1=0,
\end{align}
and
\begin{align}
	\tau_2=
	\frac{3}{2\beta_\ast^2}
	\frac{Y_\ast}{\rho_\ast^3}
	\left(4F^{\prime\prime}_\ast
	+3Y_\ast F^{\prime\prime\prime}_\ast\right).
\end{align}
The coefficient $\tau_2$ may be related to the cubic coefficient of
the background metric. From
\begin{align}
	\sigma_3
	=\frac{1}{6}\partial_\rho^3 f(\rho_\ast,\beta_\ast)=
	\frac{3Y_\ast^2}{2\rho_\ast}
	\left(4F^{\prime\prime}_\ast
	+3Y_\ast F^{\prime\prime\prime}_\ast\right),
\end{align}
one obtains
\begin{align}
	\tau_2=\frac{\rho_\ast}{\beta_\ast}\sigma_3.
\end{align}
Thus, in the large-mass expansion,
\begin{align}
	\tau_0=\frac{2}{3\beta_\ast},
	\quad
	\tau_1=0,
	\quad
	\tau_2=-\frac{5}{6\beta_\ast}
	+\left(\frac{3}{4}+\frac{5c}{3}\right)
	+\mathcal{O}(\beta_\ast).
\end{align}
The leading non-vanishing term in $\partial_\beta f$ is therefore
the constant term $\tau_0$. Hence, for a fixed polymerized theory,
\begin{align}
	k=3,\quad q=0.
\end{align}

Substitution into the general DHBI estimate gives the potentially divergent crossing contribution\footnote{For positive-energy shells and approach from the $x>0$ side, the coefficient is negative since $\sigma_3=-5/6+\mathcal{O}(\beta_\ast)<0$. Thus the crossing term drives $f_\mA\to-\infty$, so the local Misner--Sharp diagnostic $M_{\rm MS}^{(\mA)}=r(1-f_\mA)/2$ diverges to $+\infty$. This is meant as local geometric mass inflation, not as a divergence of the conserved polymerized integration constant.}
\begin{align}
	\frac{\delta_\mC\delta_\mD}{f_\mB}
	\simeq
	\left(-\frac{8}{15}\mu_\mC\mu_\mD+\mathcal{O}(\beta_\ast)\right)\frac{1}{x^3}.
\end{align}

Thus the fixed-theory perturbation does not satisfy the suppression
condition $2q-k>0$.  Instead, the DHBI crossing term diverges as
$x\to 0$. As a function of the near-horizon coordinate this
divergence is stronger than the $x^{-1}$ behavior found for the non-degenerate Hayward inner horizon. The physical time dependence, however, is different.  Because the tuned horizon has vanishing surface gravity and \(f_\mB\sim x^3\), the approach to the horizon along an outgoing null ray is power-law, $x\sim v^{-1/2}$, rather than exponential in the advanced time $v$. The amplification signaled by the fixed-theory DHBI calculation is therefore a power-law amplification, not the standard exponential mass-inflation behavior of a non-degenerate Cauchy horizon.

It is useful to contrast the fixed-theory variation relevant for the shell perturbation with a different, formal operation: a variation along a retuned family of inner-extremal geometries. In the latter case the parameters entering $\tilde f$ are allowed to vary with $\beta$, and hence with the mass parameter, so that the inner horizon remains triple degenerate at every point of the family. Let $f_{\rm ret}(\rho,\beta)$ denote the corresponding metric function. If $\rho_{-}(\beta)$ denotes the associated horizon position, then

\begin{align}
	f_{\rm ret}(\rho_{-}(\beta),\beta)=0,\quad
	\partial_\rho f_{\rm ret}(\rho_{-}(\beta),\beta)=0,
\end{align}
and
\begin{align}
    \partial_\rho^2 f_{\rm ret}(\rho_{-}(\beta),\beta)=0.
\end{align}
Differentiating the first condition with respect to \(\beta\) gives
\begin{align}
	0=\frac{d}{d\beta}f_{\rm ret}(\rho_{-}(\beta),\beta)
	={\left(\partial_\beta f_{\rm ret}\right)}_\rho
	+\rho_{-}'(\beta)\,\partial_\rho f_{\rm ret}.
\end{align}
Since $\partial_\rho f_{\rm ret}=0$ at the degenerate horizon,
\begin{align}
	\left.{\left(\partial_\beta f_{\rm ret}\right)}_\rho
	\right|_{\rho=\rho_{-}(\beta)}
	=0.
\end{align}
Differentiating the second degeneracy condition similarly yields
\begin{align}
	\left.
	\partial_\rho{\left(\partial_\beta f_{\rm ret}\right)}_\rho
	\right|_{\rho=\rho_{-}(\beta)}
	=0.
\end{align}
Therefore, when the expansion is performed around the retuned inner
horizon, $x=\rho-\rho_{-}(\beta)$, the derivative has the structure
\begin{align}
	{\left(\partial_\beta f_{\rm ret}\right)}_\rho
	=\tau^{\rm ret}_2 x^2+O(x^3).
\end{align}
Thus a retuned inner-extremal family has $q=2$.

For the explicit family used for comparison, the metric function can be
written in the factorized form
\begin{align}
	f_{\rm ret}(\rho,\beta)=(\beta\rho-1)\,\bigl(\rho-\rho_{-}(\beta)\bigr){}^3\frac{N_0+N_1\rho+N_2\rho^2}{D_6\rho^6+D_3\rho^3+D_0}, \label{eq:f(rho)}
\end{align}
where
\begin{align}
    	\rho_{-}(\beta)=1+c\beta, \quad c\neq0
\end{align}

with the coefficients $N_i$ and $D_i$ given in Appendix~\ref{sec:app}. Defining
\begin{align}
	G(\rho,\beta)
	=
	(\beta\rho-1)
	\frac{N_0+N_1\rho+N_2\rho^2}
	{D_6\rho^6+D_3\rho^3+D_0},
\end{align}
we have $f_{\rm ret}=x^3G$. Since the DHBI derivative is taken at
fixed $\rho$,
\begin{align}
	{\left(\partial_\beta f_{\rm ret}\right)}_\rho
	=
	3x^2{\left(\partial_\beta x\right)}_\rho G
	+x^3{\left(\partial_\beta G\right)}_\rho.
\end{align}
Using ${\left(\partial_\beta x\right)}_\rho=-\rho_{-}'(\beta)=-c$,
the leading term is
\begin{align}
	{\left(\partial_\beta f_{\rm ret}\right)}_\rho
	=
	-3c\,G(\rho_{-}(\beta),\beta)x^2+O(x^3),
\end{align}
so that
\begin{align}
	\tau^{\rm ret}_2=-3c\,G(\rho_{-}(\beta),\beta).
\end{align}
Substituting the explicit coefficients gives the relation for $\tau^{\rm ret}_2$ which can also be seen from Eq.~\eqref{eq:tau2gen}. In the large-mass limit this becomes
\begin{align}
	\tau^{\rm ret}_2
	=
	\frac{5}{2}c
	-\frac{3}{4}c(3+10c)\beta_\ast
	+\mathcal{O}(\beta_\ast^2).
\end{align}
Combining $q=2$ with the triple zero of the background metric,
$k=3$, the DHBI crossing term behaves as
\begin{align}
	\frac{\delta_\mC\delta_\mD}{f_\mB}
	\simeq
	-\frac{15}{2}c^2\beta_\ast^2\mu_\mC\mu_\mD\,x,
\end{align}
which vanishes as $x\to0$.

The two calculations therefore answer different questions. The retuned variation is mathematically consistent and reproduces the suppression mechanism discussed in Ref.~\cite{CRFLCV:22:deg}: if one moves along a family in which the geometry is kept inner extremal by retuning the parameters of $\tilde f$, then $\tau^{\rm ret}_0=\tau^{\rm ret}_1=0$ and the DHBI crossing term is suppressed. This is not, however, the DHBI perturbation of one fixed polymerized-vacuum theory.  In a fixed theory, the function $\tilde f$ is held fixed and the shells change
only $m$. The neighboring quadrants are then displaced away from the degenerate branch, as shown in Fig.~\ref{fig:plotfinext}, and the relevant derivative has the non-zero constant term $\tau_0=2/(3\beta_\ast)$. Within the colliding-shell model, the tuned inner-extremal polymerized background therefore exhibits a power-law amplification near the triple horizon. Establishing stability would require an additional dynamical analysis, including the backreaction of the shells, showing either that this fixed-theory amplification is cut off or that the geometry is driven toward an appropriate retuned inner-extremal configuration.

\section{Discussion}\label{sec:discussion}

We have employed the LTB collapse framework to generate static, spherically symmetric black hole solutions. A distinguished subset of these geometries consists of RBHs, which avoid the unbounded curvature scalars that typically afflict classical black hole spacetimes through the introduction of a minimal length scale, as expected from an underlying theory of quantum gravity. In our construction, these quantum-gravity–motivated corrections are encoded in the function $\tilde{f}$ appearing in Eq.~\eqref{eq:f}. The framework admits a Birkhoff-type theorem, ensuring that the resulting static configurations are unique within the class considered. Our analysis was built upon this construction, and below we summarize the main results obtained in this work.

We have briefly reviewed the consistent procedure for constructing static, spherically symmetric solutions within our framework, and discussed the Hayward model as a representative example of RBHs with a non-degenerate inner horizon. We then extended the analysis to obtain RBHs admitting a degenerate inner horizon, characterized by vanishing surface gravity. To the best of our knowledge, this constitutes the first construction of such inner-extremal RBHs in four-dimensional spacetime within a polymerized vacuum setting. Owing to the Birkhoff-type theorem satisfied by the theory, these configurations are unique once the parameters are fixed. We have shown, however, that the existence of a degenerate inner horizon requires a fine-tuning of the mass parameter with respect to the fundamental scales entering the theory. In other words, the degeneracy condition imposes a non-trivial relation between the mass and the model parameters. As a consequence, inner-extremal RBHs cannot be obtained for an arbitrary, unconstrained mass; the extremal configuration arises only on a specific submanifold of the parameter space.

RBHs possessing inner horizons are typically plagued by the mass inflation instability, whereby small perturbations near the Cauchy horizon undergo amplification, potentially driving the spacetime toward singular behavior. This represents a serious conceptual issue, since one would not wish the quantum-gravity effects responsible for regularizing the geometry to be undermined by classical instabilities. In this work, we derived a consistent and general criterion for the occurrence of mass inflation, summarized in Table~\ref{table:mass-inflation}. We then applied this criterion to two representative cases: the Hayward model with a non-degenerate inner horizon, and the inner-extremal configuration constructed within our polymerized vacuum framework. 

Our analysis shows that the inner-extremal case is more subtle than the non-degenerate Hayward geometry. For a formal variation along a retuned family of inner-extremal solutions, the DHBI criterion gives a suppressed crossing term, in agreement with the expected effect of the degenerate inner horizon. However, for a fixed polymerized-vacuum theory, the null shells change the integration constant $m$ while the function $\tilde{f}$ is held fixed; the neighboring regions are then displaced away from the degenerate branch, and the colliding-shell model instead yields a power-law amplification near the triple horizon. 

The Hayward model, by contrast, exhibits the usual divergent DHBI term associated with a non-degenerate inner horizon. Importantly, however, this divergence becomes significant only when the shell collision occurs at an areal-radius separation from the inner horizon that lies below the Planck scale. In this regime, the classical effective spacetime description used in our analysis is no longer reliable. By relating the regularization scale $\ell$ to the LQG area gap, and assuming macroscopic black hole masses, we find that the onset of the Hayward mass-inflation effect is shifted outside the expected domain of validity of the effective geometry.

There are several important directions in which the present construction can be extended. Although we have developed an explicit framework based on LTB collapse, our analysis has been restricted to static vacuum regions and to an idealized colliding null-shell model. The DHBI calculation should therefore be understood as a near-horizon diagnostic of blueshift-driven amplification, not as a substitute for a full dynamical solution of the perturbed polymerized theory. Establishing stability would require evolving the metric, the matter or null-shell perturbations, and the reference-clock sector within the same fixed effective theory, and monitoring invariant quantities such as curvature scalars or the canonical curvature and extrinsic-curvature variables. Such an analysis is needed to determine whether the amplification found in the fixed-theory inner-extremal case is dynamically cut off, converted into a pathology of the clock foliation, or develops into a genuine curvature singularity. Semiclassical effects, including vacuum polarization and quantum stress-energy fluxes, may further modify the inner-horizon dynamics, as observed in Ref.~\cite{Mc:23}. Determining whether the LQG-motivated minimal length scale can consistently control inner-horizon instabilities once quantum fields are incorporated therefore remains a key open problem.

\section*{acknowledgments}
This work is supported by the National Natural Science Foundation of China (Grant No. 12505081) and the start-up funding from Westlake University. I.S. is supported by the Institute for Theoretical Sciences at Westlake University.

\appendix

\begin{widetext}
\section{Inner-extremal RBH parameters}\label{sec:app}

Requiring that the metric function defined by Eq.~\eqref{eq:f-deg-initial} admits four roots $r_i$ with $i=1,2,3,4$ enables us to express the mass $m$ and the parameters $a_2$, $b_1$, and $b_2$ in terms of these roots. The resulting expressions are as follows:

\begin{equation}
\begin{split}
m=\frac{1}{2}\Bigl(&
r_1^3 \bigl(r_3 r_4+r_2(r_3+r_4)\bigr)
+r_1^2(r_2+r_3+r_4)\bigl(r_3 r_4+r_2(r_3+r_4)\bigr) \\
&+r_1\bigl(r_3 r_4+r_2(r_3+r_4)\bigr)
\bigl(r_2^2+r_3^2+r_3r_4+r_4^2+r_2(r_3+r_4)\bigr) \\
&+r_2r_3r_4
\bigl(r_2^2+r_3^2+r_3r_4+r_4^2+r_2(r_3+r_4)\bigr)
\Bigr) \\
&\times
\Bigl(
r_2r_3r_4(r_2+r_3+r_4)
+r_1^2\bigl(r_3r_4+r_2(r_3+r_4)\bigr) \\
&\qquad
+r_1(r_2+r_3+r_4)\bigl(r_3r_4+r_2(r_3+r_4)\bigr)
\Bigr)^{-1}.
\end{split}
\end{equation}

	\begin{equation}
		\begin{split}
			a_2 = &\Bigl( r_2 r_3 r_4 (r_2 + r_3 + r_4) + r_1^2 (r_3 r_4 + r_2 (r_3 + r_4)) + r_1 (r_2 + r_3 + r_4) (r_3 r_4 + r_2 (r_3 + r_4)) \Bigr) \\
			&\times \Bigl( r_2^2 r_3^2 r_4^2 (r_3 r_4 + r_2 (r_3 + r_4))+ r_1 r_2 r_3 r_4 (r_3 r_4 + r_2 (r_3 + r_4))^2 \\
			&\quad + r_1^2 (r_3 r_4 + r_2 (r_3 + r_4)) \bigl( r_3^2 r_4^2 + r_2 r_3 r_4 (r_3 + r_4) + r_2^2 (r_3^2 + r_3 r_4 + r_4^2) \bigr) \\
			&\quad + r_1^3 \bigl( r_3^2 r_4^2 (r_3 + r_4) + r_2 r_3 r_4 (r_3 + r_4)^2 + r_2^3 (r_3^2 + r_3 r_4 + r_4^2) \\
			&\qquad + r_2^2 (r_3 + r_4) (r_3^2 + r_3 r_4 + r_4^2) \bigr) \Bigr) \\
			&\Big/ \Bigl[ \Bigl( r_1^3 (r_3 r_4 + r_2 (r_3 + r_4)) + r_1^2 (r_2 + r_3 + r_4) (r_3 r_4 + r_2 (r_3 + r_4)) \\
			&\quad + r_2 r_3 r_4 \bigl( r_2^2 + r_3^2 + r_3 r_4 + r_4^2 + r_2 (r_3 + r_4) \bigr) \\
			&\quad + r_1 (r_3 r_4 + r_2 (r_3 + r_4)) \bigl( r_2^2 + r_3^2 + r_3 r_4 + r_4^2 + r_2 (r_3 + r_4) \bigr) \Bigr)^2 \Bigr],
		\end{split}
	\end{equation}
	
	\begin{equation}
		\begin{split}
			b_1 = &\Bigl( r_1^2 (r_3 r_4 + r_2 (r_3 + r_4)) \bigl( r_2^2 (r_3 + r_4) + r_3 r_4 (r_3 + r_4) + r_2 (r_3^2 + r_3 r_4 + r_4^2) \bigr) \\
			&\quad + r_2 r_3 r_4 \bigl( r_3^2 r_4^2 + r_2 r_3 r_4 (r_3 + r_4) + r_2^2 (r_3^2 + r_3 r_4 + r_4^2) \bigr) \\
			&\quad + r_1 (r_3 r_4 + r_2 (r_3 + r_4)) \bigl( r_3^2 r_4^2 + r_2 r_3 r_4 (r_3 + r_4) + r_2^2 (r_3^2 + r_3 r_4 + r_4^2) \bigr) \\
			&\quad + r_1^3 \bigl( r_2^3 (r_3 + r_4) + r_2^2 (r_3 + r_4)^2 + r_3 r_4 (r_3^2 + r_3 r_4 + r_4^2) \\
			&\qquad + r_2 (r_3 + r_4) (r_3^2 + r_3 r_4 + r_4^2) \bigr) \Bigr) \\
			&\Big/ \Bigl( r_1^3 (r_3 r_4 + r_2 (r_3 + r_4)) + r_1^2 (r_2 + r_3 + r_4) (r_3 r_4 + r_2 (r_3 + r_4)) \\
			&\quad + r_2 r_3 r_4 \bigl( r_2^2 + r_3^2 + r_3 r_4 + r_4^2 + r_2 (r_3 + r_4) \bigr) \\
			&\quad + r_1 (r_3 r_4 + r_2 (r_3 + r_4)) \bigl( r_2^2 + r_3^2 + r_3 r_4 + r_4^2 + r_2 (r_3 + r_4) \bigr) \Bigr),
		\end{split}
	\end{equation}
	
	\begin{equation}
		\begin{split}
			b_2 = &\Bigl( r_1^2 r_2^2 r_3^2 r_4^2 \bigl( r_3 r_4 + r_2 (r_3 + r_4) + r_1 (r_2 + r_3 + r_4) \bigr) \\
			&\quad \times \bigl( r_2 r_3 r_4 (r_2 + r_3 + r_4) + r_1^2 (r_3 r_4 + r_2 (r_3 + r_4)) \\
			&\qquad + r_1 (r_2 + r_3 + r_4) (r_3 r_4 + r_2 (r_3 + r_4)) \bigr) \Bigr) \\
			&\Big/ \Bigl[ \Bigl( r_1^3 (r_3 r_4 + r_2 (r_3 + r_4)) + r_1^2 (r_2 + r_3 + r_4) (r_3 r_4 + r_2 (r_3 + r_4)) \\
			&\quad + r_2 r_3 r_4 \bigl( r_2^2 + r_3^2 + r_3 r_4 + r_4^2 + r_2 (r_3 + r_4) \bigr) \\
			&\quad + r_1 (r_3 r_4 + r_2 (r_3 + r_4)) \bigl( r_2^2 + r_3^2 + r_3 r_4 + r_4^2 + r_2 (r_3 + r_4) \bigr) \Bigr)^2 \Bigr].
		\end{split}
	\end{equation}

When expressing the metric function of the inner-extremal model in Eq.~\eqref{eq:f(rho)} in terms of dimensionless parameters, we introduce the following function in order to write it in a more compact form.

\begin{align}
N_0 = 1 + \beta + \beta c \bigl(2 + \beta (3 + c + \beta c (3 + \beta c)) \bigr),
\end{align}

\begin{align}
	N_1 = 3 + \beta \bigl(4 + \beta + 3c + \beta c (8 + \beta (3 + c (4 + \beta (3 + \beta c)))) \bigr),
\end{align}

\begin{align}
N_2 = 1 + \beta (1 + \beta c) (3 + \beta + \beta^2 c),
\end{align}

\begin{align}
	D_0 = (1 + \beta c)^5 (1 + \beta + \beta^2 c),
\end{align}

\begin{align}
D_3 = (1 + \beta c)^2 \bigl(5 + \beta (1 + \beta c) (7 + 2 \beta (1 + \beta c) (3 + \beta + \beta^2 c)) \bigr),
\end{align}

\begin{align}
D_6 = \beta \bigl(1 + \beta (1 + \beta c) (3 + \beta + \beta^2 c) \bigr).
\end{align}

Finally, we present the expressions corresponding to the derivative of the retuned inner-extremal metric function with respect to the parameter $\beta$, together with its leading contribution in the expansion around the inner horizon. This expansion is performed in the large-mass limit and is given by

	\begin{align}
		\partial_\beta 	f_{\rm ret}=-\frac{c (-1 + \beta + \beta^2 c) \bigl(5 + 2 \beta (1 + \beta c) (4 + \beta + \beta^2 c)\bigr) }{(1 + \beta c)^3 (2 + \beta + \beta^2 c) \bigl(1 + \beta (1 + \beta c) (1 + \beta + \beta^2 c)\bigr)}x^2+\mathcal{O}(x^3). \label{eq:tau2gen}
	\end{align}
\end{widetext}


\begin{thebibliography}{100}
	
	\bibitem{H:76} S. W. Hawking,
	Breakdown of predictability in gravitational collapse,
	\href{https://doi.org/10.1103/PhysRevD.14.2460}{Phys. Rev. D \textbf{14}, 2460 (1976)}.
	
	\bibitem{CP:19} V.\ Cardoso and P.\ Pani,
	Testing the nature of dark compact objects: a status report,
	{\href{https://doi.org/10.1007/s41114-019-0020-4}{ {Living Rev.\ Relativ.} \textbf{22}, 4  (2019)}}.
	
	\bibitem{BCNS:19} L.\ Barack, V.\ Cardoso, S.\ Nissanke, and T.\ P.\ Sotiriou (eds.),
	Black holes, gravitational waves and fundamental physics: a roadmap,
	{\href{https://doi.org/10.1088/1361-6382/ab0587}{Class.\ Quantum Gravity \textbf{36}, 143001 (2019)}}.
	
	\bibitem{M:23} S.\ Murk,
	Nomen non est omen: Why it is too soon to identify ultra-compact objects as black holes,
	\href{https://doi.org/10.1142/S0218271823420129}{Int.\ J.\ Mod.\ Phys.\ D \textbf{32}, 2342012 (2023)}.
	
	\bibitem{B:68} J.\ M.\ Bardeen 
	Non-singular general relativistic gravitational collapse,
	\textit{Proceedings of the International Conference GR5} (Tbilisi University Press, Tbilisi, 1968).
	
	\bibitem{D:92} I.\ Dymnikova, 
	Vacuum nonsingular black hole,
	\href{https://doi.org/10.1007/BF00760226}{Gen.\ Relativ.\ Gravit.\ \textbf{24}, 235 (1992)}.
	
	\bibitem{H:06} S.\ A.\ Hayward,
	Formation and Evaporation of Nonsingular Black Holes,
	\href{https://doi.org/10.1103/PhysRevLett.96.031103}{Phys.\ Rev.\ Lett.\ \textbf{96}, 031103 (2006)}.
	
	\bibitem{MM:04} P.\ O.\ Mazur and E.\ Mottola,
	Gravitational vacuum condensate stars,
	\href{https://doi.org/10.1073/pnas.0402717101}{Proc.\ Natl.\ Acad.\ Sci.\ \textbf{101}, 9545 (2004)}.
	
	\bibitem{MM:23-grav} P.\ O.\ Mazur and E.\ Mottola,
	Gravitational Condensate Stars: An Alternative to Black Holes,
	\href{https://doi.org/10.3390/universe9020088}{Universe \textbf{9}, 88 (2023)}.
	
	\bibitem{E:73} H.\ G.\ Ellis,
	Ether flow through a drainhole: A particle model in general relativity,
	\href{https://doi.org/10.1063/1.1666161}{J.\ Math.\ Phys.\ \textbf{14}, 104 (1973)}.
	
	\bibitem{MT:88} M.\ S.\ Morris and K.\ S.\ Thorne,
	Wormholes in spacetime and their use for interstellar travel: A tool for teaching general relativity,
	\href{https://doi.org/10.1119/1.15620}{Am.\ J.\ Phys.\ \textbf{56}, 395 (1988)}.
	
	\bibitem{SV:19} A.\ Simpson and M.\ Visser,
	Black-bounce to traversable wormhole,
	\href{https://doi.org/10.1088/1475-7516/2019/02/042}{J.\ Cosmol.\ Astropart.\ Phys.\ 02 (2019) 042}.
	
	\bibitem{LM:02} O.\ Lunin and S.\ D.\ Mathur,
	AdS/CFT duality and the black hole information paradox,
	\href{https://doi.org/10.1016/S0550-3213(01)00620-4}{Nucl.\ Phys.\ B \textbf{623}, 342, (2002)}.
	
	\bibitem{M:05} S.\ D.\ Mathur,
	The fuzzball proposal for black holes: an elementary review,
	\href{https://doi.org/10.1002/prop.200410203}{Fortsch.\ Phys.\  \textbf{53}, 793 (2005)}.
	
	\bibitem{PI:89} E. Poisson and W. Israel, 
	Inner-horizon instability and mass inflation in black holes,
	\href{https://doi.org/10.1103/PhysRevLett.63.1663}{Phys. Rev. Lett. \textbf{63}, 1663 (1989).}
	
	\bibitem{PI:90} E. Poisson and W. Israel, 
	Internal structure of black holes,
	\href{https://doi.org/10.1103/PhysRevD.41.1796}{Phys. Rev. D \textbf{41}, 1796 (1990).}
	
	\bibitem{BS:95} P. R. Brady and J. D. Smith,
	Black Hole Singularities: A Numerical Approach,
	\href{https://doi.org/10.1103/PhysRevLett.75.1256}{Phys. Rev. Lett. \textbf{75}, 1256 (1995).}
	
	\bibitem{HA:10} A. J. S. Hamilton and P. P. Avelino, 
	The physics of the relativistic counter-streaming instability that drives mass inflation inside black holes,
	\href{https://doi.org/10.1016/j.physrep.2010.06.002}{Phys. Rept. \textbf{495}, 1 (2010).}
	
	\bibitem{HY:11} D.-i. Hwang and D.-h. Yeom,
	Internal structure of charged black holes,
	\href{https://doi.org/10.1103/PhysRevD.84.064020}{Phys. Rev. D \textbf{84}, 064020 (2011).}
	
	\bibitem{CRFLV:24} R. Carballo-Rubio, F. Di Filippo, S. Liberati, and M. Visser,
	Mass Inflation without Cauchy Horizons,
	\href{https://doi.org/10.1103/PhysRevLett.133.181402}{Phys. Rev. Lett. \textbf{133}, 181402 (2024).}
	
	\bibitem{FZ:26} V. P. Frolov and A. Zelnikov,
	Regular Black Holes in Quasitopological Gravity: Null Shells and Mass Inflation,
	\href{https://doi.org/10.48550/arXiv.2601.01861}{arXiv: 2601.01861 [gr-qc] (2026).}
	
	\bibitem{FZ:17} V. P. Frolov and A. Zelnikov,
	Quantum radiation from an evaporating nonsingular black hole,
	\href{https://doi.org/10.1103/PhysRevD.95.124028}{Phys. Rev. D 95, 124028 (2017).}
	
	\bibitem{BMM:11} E. G. Brown, R. B. Mann, and L. Modesto,
	Mass inflation in the loop black hole,
	\href{https://doi.org/10.1103/PhysRevD.84.104041}{Phys. Rev. D \textbf{84}, 104041 (2011)}
	
	\bibitem{BKS:21} A. Bonanno A.-P. Khosravi, and F. Saueressig,
	Regular black holes with stable cores,
	\href{https://doi.org/10.1103/PhysRevD.103.124027}{Phys. Rev. D \textbf{103}, 124027 (2021).}
	
	\bibitem{HKSWE:22} V. Husain, J. G. Kelly, R. Santacruz, and E. Wilson-Ewing,
	Quantum gravity of dust collapse: shock waves from black holes,
	\href{https://doi.org/10.1103/PhysRevLett.128.121301}{Phys. Rev. Lett. \textbf{128}, 121301 (2022).}
	
	\bibitem{CRFLPV:22} R. Carballo-Rubio, F. Di Filippo, S. Liberati, C. Pacilio, and M. Visser,
	On the Inner Horizon Instability of Non-Singular Black Holes,
	\href{https://doi.org/10.3390/universe8040204}{Universe \textbf{2022}, 8(4), 204 (2022).}
	
	\bibitem{BRSZ:21} M.l Bertipagani, M. Rinaldi, L. Sebastiani, and S. Zerbini,
	Non-singular black holes and mass inflation in modified gravity,
	\href{https://doi.org/10.1016/j.dark.2021.100853}{Physics of the Dark Universe \textbf{33}, 100853 (2021).}
	
	\bibitem{BF:22} A. Bonanno and F. Saueressig,
	Stability properties of Regular Black Holes,
	\href{https://doi.org/10.48550/arXiv.2211.09192}{arXiv: 2211.09192 [gr-qc] (2022).}
	
	\bibitem{Bambi:book:23} C.\ Bambi (ed.),
	\href{https://doi.org/10.1007/978-981-99-1596-5}{\textit{Regular Black Holes: Towards a New Paradigm of Gravitational Collapse} (Springer Singapore, 2023)}.
	
	\bibitem{BBCRG:22} C. Barceló, V. Boyanov, R. Carballo-Rubio, and L. J. Garay,
	Classical mass inflation versus semiclassical inner horizon inflation,
	\href{https://doi.org/10.1103/PhysRevD.106.124006}{Phys. Rev. D \textbf{106}, 124006 (2022).}
	
	\bibitem{Mc:23} T. McMaken,
	Semiclassical instability of inner-extremal regular black holes,
	\href{ https://doi.org/10.1103/PhysRevD.107.125023}{Phys. Rev. D \textbf{107}, 125023 (2023).}
	
	\bibitem{BPS:25} A. Bonanno, A. Panassiti, and F. Saueressig,
	Cauchy Horizon (In)Stability of Regular Black Holes,
	\href{https://doi.org/10.48550/arXiv.2507.03581}{arXiv: 2507.03581 [gr-qc] (2025).}
	
	\bibitem{CLWZ:24} L.-M. Cao, L.-Y. Li, L.-B. Wu, and Y.-S. Zhou, 
	The instability of the inner horizon of the quantum-corrected black hole,
	\href{https://doi.org/10.1140/epjc/s10052-024-12832-4}{Eur. Phys. J. C \textbf{84}, 507 (2024). }
	
	\bibitem{CRFLCV:22:deg} R. Carballo-Rubio, F. Di Filippo, S. Liberati, C. Pacilio, and M. Visser,
	Regular black holes without mass inflation instability,
	\href{https://doi.org/10.1007/JHEP09(2022)118}{ J. High Energ. Phys. \textbf{2022}, 118 (2022).}
	
	\bibitem{pt:69} R.\ Pellicer and R.\ J.\ Torrence,
	Nonlinear Electrodynamics and General Relativity,
	\href{https://doi.org/10.1063/1.1665019}{J.\ Math.\ Phys.\ \textbf{10}, 1718 (1969)}.
	
	\bibitem{bmss:79} K.\ A.\ Bronnikov, V.\ N.\ Melnikov, G.\ N.\ Shikin, and K.\ P.\ Staniukovich,
	Scalar, electromagnetic, and gravitational fields interaction: Particlelike solutions,
	\href{https://doi.org/10.1016/0003-4916(79)90235-5}{Ann.\ Phys.\ \textbf{118}, 84 (1979)}.
	
	\bibitem{ag:98} E.\ Ayón-Beato and E.\ García,
	Regular Black Hole in General Relativity Coupled to Nonlinear Electrodynamics,	
	\href{https://doi.org/10.1103/PhysRevLett.80.5056}{Phys.\ Rev.\ Lett.\ \textbf{80}, 5056 (1998)}.
	
	\bibitem{ag:99a} E.\ Ayón-Beato and E.\ García,
	Non-Singular Charged Black Hole Solution for Non-Linear Source,
	\href{https://doi.org/10.1023/A:1026640911319}{Gen.\ Relativ.\ Gravit.\ \textbf{31}, 629 (1999)}.
	
	\bibitem{ag:99b} E.\ Ayón-Beato and E.\ García,
	New regular black hole solution from nonlinear electrodynamics,
	\href{https://doi.org/10.1016/S0370-2693(99)01038-2}{Phys.\ Lett.\ B \textbf{464}, 25 (1999)}.
	
	\bibitem{b:00} K.\ A.\ Bronnikov,
	Comment on “Regular Black Hole in General Relativity Coupled to Nonlinear Electrodynamics”,
	\href{https://doi.org/10.1103/PhysRevLett.85.4641}{Phys.\ Rev.\ Lett.\ \textbf{85}, 4641 (2000)}.
	
	\bibitem{b:01} K.\ A.\ Bronnikov,
	Regular magnetic black holes and monopoles from nonlinear electrodynamics,
	\href{https://doi.org/10.1103/PhysRevD.63.044005}{Phys.\ Rev.\ D \textbf{63}, 044005 (2001)}.
	
	\bibitem{bh:02} A.\ Burinskii and S.\ R.\ Hildebrandt,
	New type of regular black holes and particlelike solutions from nonlinear electrodynamics,
	\href{https://doi.org/10.1103/PhysRevD.65.104017}{Phys.\ Rev.\ D \textbf{65}, 104017 (2002)}.
	
	\bibitem{d:04} I.\ Dymnikova,
	Regular electrically charged vacuum structures with de Sitter centre in nonlinear electrodynamics coupled to general relativity,
	\href{https://doi.org/10.1088/0264-9381/21/18/009}{Class.\ Quantum Gravity \textbf{21}, 4417 (2004)}.
	
	\bibitem{bv:14} L.\ Balart and E.\ C.\ Vagenas,
	Regular black holes with a nonlinear electrodynamics source,
	\href{https://doi.org/10.1103/PhysRevD.90.124045}{Phys.\ Rev.\ D \textbf{90}, 124045 (2014)}.
	
	\bibitem{fw:16} Z.-Y.\ Fan and X.\ Wang,
	Construction of regular black holes in general relativity,
	\href{https://doi.org/10.1103/PhysRevD.94.124027}{Phys. Rev. D \textbf{94}, 124027 (2016)}.
	
	\bibitem{b:17} K.\ A.\ Bronnikov,
	Comment on “Construction of regular black holes in general relativity”,
	\href{https://doi.org/10.1103/PhysRevD.96.128501}{Phys.\ Rev.\ D \textbf{96}, 128501 (2017)}.
	
	\bibitem{tsa:18} B.\ Toshmatov, Z.\ Stuchlík, and B.\ Ahmedov,
	Comment on “Construction of regular black holes in general relativity”,
	\href{https://doi.org/10.1103/PhysRevD.98.028501}{Phys.\ Rev.\ D \textbf{98}, 028501 (2018)}.
	
	\bibitem{b:23} K.\ A.\ Bronnikov, 
	Regular Black Holes Sourced by Nonlinear Electrodynamics,
	in C.\ Bambi (ed.) \href{https://doi.org/10.1007/978-981-99-1596-5}{Regular Black Holes: Towards a New Paradigm of Gravitational Collapse (Springer Singapore, 2023)}.
	
	\bibitem{BCH:25} P. Bueno, P. A. Cano, and R. A. Hennigar,
	Regular black holes from pure gravity,
	\href{https://doi.org/10.1016/j.physletb.2025.139260}{Phys.\ Lett.\ B \textbf{861}, 139 (2025)}.
	
	\bibitem{BCHM:25a} P. Bueno, P. A. Cano, R. A. Hennigar, and Á. J. Murcia
	Regular black holes from thin-shell collapse,
	\href{https://doi.org/10.1103/PhysRevD.111.104009}{Phys. Rev. D \textbf{111}, 104009 (2025)}.
	
	\bibitem{BCHM:25b} P. Bueno, P. A. Cano, R. A. Hennigar, and Á. J. Murcia
	Dynamical Formation of Regular Black Holes,
	\href{https://doi.org/10.1103/PhysRevLett.134.181401}{Phys. Rev. Lett. \textbf{134}, 181401 (2025)}.
	
	\bibitem{BCHMC:25} P. Bueno, P. A. Cano, R. A. Hennigar, Á. J. Murcia, and A. Vincente-Cano,
	Regular black holes from Oppenheimer-Snyder collapse,
	\href{https://doi.org/10.1103/qrbb-mdvm}{Phys. Rev. D \textbf{112}, 064039 (2025)}.
	
	\bibitem{HKMS:25} R.\ A.\ Hennigar, D.\ Kubizňák, S.\ Murk, and I.\ Soranidis,
	Thermodynamics of regular black holes in anti-de Sitter space, \href{https://doi.org/10.1007/JHEP11(2025)121}{J. High Energ. Phys. 2025, 121 (2025)}.
	
	\bibitem{FKSZ:25} V. P. Frolov, A. Koek, J. P. Soto, and A. Zelnikov,
	Regular black holes inspired by quasitopological gravity,
	\href{https://doi.org/10.1103/PhysRevD.111.044034}{Phys. Rev. D \textbf{111}, 044034 (2025).}
	
	\bibitem{F:25} P. G. S. Fernandes,
	Singularity resolution and inflation from an infinite tower of regularized curvature corrections,
	\href{https://doi.org/10.1103/763p-htct}{Phys. Rev. D \textbf{112}, 084028  (2025)}.
	
	\bibitem{BCR:26} J. Borissova and R. Carballo-Rubio,
	Regular black holes from pure gravity in four dimensions,
	\href{https://doi.org/10.48550/arXiv.2602.16773}{arXiv: 2602.16773 [gr-qc] (2026).}
	
	\bibitem{BCHM:26} P. Bueno, P. A. Cano, R. A. Hennigar, and Á. J. Murcia,
	Regular black hole formation in four-dimensional nonpolynomial gravities,
	\href{https://doi.org/10.1103/8f3j-zcxh}{Phys. Rev. D \textbf{113}, 024019 (2026).}
	
	\bibitem{GLRSW:25} K. Giesel, H. Liu, E. Rullit, P. Singh, and S. A. Weigl,
	Embedding generalized Lemaître-Tolman-Bondi models in polymerized spherically symmetric spacetimes,
	\href{https://doi.org/10.1103/PhysRevD.110.104017}{Phys. Rev. D \textbf{110}, 104017 (2024).}
	
	\bibitem{GLSW:25} K.Giesel, H. Liu, P. Singh, and S. A. Weigl,
	Regular black holes and their relationship to polymerized models and mimetic gravity,
	\href{https://doi.org/10.1103/PhysRevD.111.064064}{Phys. Rev. D \textbf{111}, 064064 (2025)}.
	
	\bibitem{GL:25} K. Giesel and H. Liu,
	From Principles to Effective Models: A Constructive Framework for Effective Covariant Actions with a Unique Vacuum Solution,
	\href{https://doi.org/10.48550/arXiv.2512.24960}{arXiv: 2512.24960 (2025)}.
	
	\bibitem{Han:2022rsx} M. Han and H. Liu,
	Covariant {\ensuremath{\mu}}{\textasciimacron}-scheme effective dynamics, mimetic gravity, and nonsingular black holes: Applications to spherically symmetric quantum gravity,
	\href{https://doi.org/10.1103/PhysRevD.109.084033}{Phys. Rev. D \textbf{109}, 084033 (2024).}
	
	\bibitem{Han:2020uhb} M. Han and H. Liu,
	Improved effective dynamics of loop-quantum-gravity black hole and Nariai limit,
	\href{https://doi.org/10.1088/1361-6382/ac44a0}{Classical Quantum Gravity \textbf{39}, 035011 (2022).}
	
%	\bibitem{BenAchour:2017ivq}
%	J.~Ben Achour, F.~Lamy, H.~Liu and K.~Noui,
%	Non-singular black holes and the Limiting Curvature Mechanism: A Hamiltonian perspective,
%	JCAP \textbf{05} (2018), 072
%	\href{https://doi.org/10.1088/1475-7516/2018/05/072}{J. Cosmol. Astropart. Phys. \textbf{05}, 072 (2018)}.
	
	\bibitem{BenAchour:2018khr} J. Ben Achour, F. Lamy, H. Liu, and K. Noui,
	Polymer Schwarzschild black hole: An effective metric,
	\href{https://doi.org/10.1209/0295-5075/123/20006}{Europhysics letters \textbf{123}, 20006 (2018).}
	
	\bibitem{B:26} J. Borissova,
	All 2D generalized dilaton theories from $d\geqslant4$ gravities,
	\href{https://doi.org/10.1103/v3j1-j6jg}{Phys. Rev. D 113, 124088 (2026).}
	
	\bibitem{BM:26} J. Borissova and J. Magueijo,
	Modified Friedmann equations and non-singular cosmologies in $d=4$ non-polynomial quasi-topological gravities,
	\href{https://doi.org/10.1088/1475-7516/2026/07/045}{J.\ Cosmol.\ Astropart.\ Phys.\ 07 (2026) 045}.
	
	\bibitem{LS:26} H. Liu and I. Soranidis,
	Regular ultracompact objects with anti-de Sitter cores as polymerized vacuum solutions,
	\href{https://doi.org/10.48550/arXiv.2604.27880}{arXiv: 2604.27880 [gr-qc] (2026).}
	
	\bibitem{BS:26} S. Bakhoda and I. Soranidis,
	Thermodynamics of polymerized vacuum regular black holes in anti-de Sitter spacetime,
	\href{https://doi.org/10.48550/arXiv.2606.13167}{arXiv:2606.13167 [gr-qc] (2026)}.
	
	\bibitem{BCH:19} P. Bueno, P. A. Cano, and R. A. Hennigar,
	(Generalized) quasi-topological gravities at all orders
	\href{https://doi.org/10.1088/1361-6382/ab5410}{Classical Quantum Gravity \textbf{37}, 015002 (2019)}.
	
	\bibitem{BCHLM:22} P. Bueno, P. A. Cano, R. A. Hennigar, M. Lu, and J. Moreno,
	Generalized quasi-topological gravities: the whole shebang,
	\href{https://doi.org/10.1088/1361-6382/aca236}{Classical Quantum Gravity \textbf{40}, 015004 (2022)}.
	
	\bibitem{MM:23} J. Moreno and Á. J. Murcia,
	Classification of generalized quasitopological gravities,
	\href{https://doi.org/10.1103/PhysRevD.108.044016}{Phys. Rev. D \textbf{108}, 044016 (2023)}.
	
	\bibitem{BHM:25} P. Bueno, R. A. Hennigar, and  Á. J. Murcia,
	Birkhoff implies Quasi-topological,
	\href{https://doi.org/10.48550/arXiv.2510.25823}{arXiv: 2510.25823 (2025)}.
	
	\bibitem{FKK:25} F. Di Filippo, I. Kolář, and D. Kubizňák,
	Inner-extremal regular black holes from pure gravity,
	\href{https://doi.org/10.1103/PhysRevD.111.L041505}{Phys. Rev. D \textbf{111}, L041505 (2025).}
	
	\bibitem{LLB:07} P. Lasky, A. Lun, and R. Burston,
	Initial value formalism for dust collapse,
	\href{https://doi.org/10.48550/arXiv.gr-qc/0606003}{arXiv: gr-qc/0606003 (2007).}
	
	\bibitem{TK:17} K. Takahashi and T. Kobayashi,
	Extended mimetic gravity: Hamiltonian analysis and gradient instabilities,
	\href{https://doi.org/10.1088/1475-7516/2017/11/038}{J. Cosmol. Astropart. Phys. \textbf{11}, 038 (2017)}.
	
	\bibitem{LMNV:19} D. Langlois, M. Mancarella, K. Noui and F. Vernizzi,
	Mimetic gravity as DHOST theories,
	\href{https://doi.org/10.1088/1475-7516/2019/02/036}{J. Cosmol. Astropart. Phys. \textbf{02}, 036 (2019)}.
	
	\bibitem{ALNV:25} J. Arrechea, S. Liberati, H. Neshat, and V. Vellucci,
	From de Sitter to anti–de Sitter singularity regularization: Theory and phenomenology,
	\href{https://doi.org/10.1103/919y-m4yj}{Phys. Rev. D \textbf{112}, 124029  (2025).}
	
	\bibitem{MS:23} S. Murk and I. Soranidis, 
	Kinematic and energy properties of dynamical regular black holes,
	\href{https://doi.org/10.1103/PhysRevD.108.124007}{Phys. Rev. D \textbf{108}, 124007 (2023).}
	
	\bibitem{DH:85} T. Dray and G. ’t Hooft,
	The effect of spherical shells of matter on the Schwarzschild black hole,
	\href{https://doi.org/10.1007/BF01215912}{Commun. Math. Phys. \textbf{99}, 613 (1985).}
	
	\bibitem{BI:91} C. Barrabès and W. Israel,
	Thin shells in general relativity and cosmology: The lightlike limit,
	\href{https://doi.org/10.1103/PhysRevD.43.1129}{Phys. Rev. D \textbf{43}, 1129 (1991).}
	
	\bibitem{DKS:26} F. Di Filippo, D.\ Kubizňák, and A. Srinivasan,
	On mass inflation and thin shells in quasi-topological gravity,
	\href{https://doi.org/10.48550/arXiv.2604.27980}{arXiv: 2604.27980 [gr-qc] (2026).}
	
	\bibitem{Rov:10} C. Rovelli
	{\href{https://doi.org/10.1017/CBO9780511755804}{\textit{Quantum Gravity} (Cambridge University Press, Cambridge, England, 2010)}}.
	
	\bibitem{RV:15} C. Rovelli and F. Vidotto,
	{\href{https://doi.org/10.1017/CBO9781107706910}{\textit{Covariant Loop Quantum Gravity: An Elementary Introduction to Quantum Gravity and Spinfoam Theory} (Cambridge University Press, Cambridge, England, 2015)}}.
	
	\bibitem{T:10} T. Thiemann,
	{\href{https://doi.org/10.1017/CBO9780511755682}{\textit{Modern Canonical Quantum General Relativity} (Cambridge University Press, Cambridge, England, 2010)}}.
	
	\bibitem{GLSW:24} K. Giesel, H. Liu, P. Singh, and S. A. Weigl,
	Generalized analysis of a dust collapse in effective loop quantum gravity: Fate of shocks and covariance,
	\href{https://doi.org/10.1103/PhysRevD.110.104016}{Phys. Rev. D \textbf{110}, 104016 (2024).}
	
	\bibitem{Giesel:2022rxi}
	K.~Giesel, M.~Han, B.~F.~Li, H.~Liu and P.~Singh,
	Spherical symmetric gravitational collapse of a dust cloud: Polymerized dynamics in reduced phase space,
	\href{https://doi.org/10.1103/PhysRevD.107.044047}{Phys. Rev. D \textbf{107}, 044047 (2023)}.
	
\end{thebibliography}
\end{document}